\documentclass[paper=a4,fontsize=11pt,DIV=18]{scrartcl}
\pdfoutput=1

\usepackage[utf8]{inputenc}		
\usepackage[T1]{fontenc}		
\usepackage{lmodern}

\usepackage[ngerman, american]{babel}							
\usepackage[autostyle]{csquotes}		

\usepackage[footnotesize]{caption}[2008/08/24]		

\usepackage{amsmath,amsfonts,amssymb,mathtools}

\usepackage{braket}
\usepackage[separate-uncertainty=true]{siunitx}
\usepackage{graphicx}
\usepackage{epstopdf}

\usepackage{color}
\usepackage[svgnames,dvipsnames]{xcolor}

\usepackage[numbers,sort&compress]{natbib}

\usepackage[colorlinks=true, pdfborder={0 0 0}, plainpages=false, citecolor=teal, linkcolor=black]{hyperref} 

\addto\captionsamerican{

}

\usepackage{authblk}

\usepackage{tikz}
\usepackage{helvet,times}
\usepackage{bm,textcomp}

\usepackage{braket}

\usepackage{pdfpages}

\hypersetup{
	pdftitle={Anti-margination of microparticles and platelets in the vicinity of branching vessels},
	pdfauthor={C. B\"acher, A. Kihm, L. Schrack, L. Kaestner, M.W. Laschke, C. Wagner, and S. Gekle},
	pdfkeywords={blood flow; margination; targeted drug delivery; blood network}
}

\begin{document}

\setcounter{page}{1} 

\title{Anti-margination of microparticles and platelets in the vicinity of branching vessels}
\date{}
\author[1]{C. B\"acher}
\author[2]{A. Kihm}
\author[1,3]{L. Schrack}
\author[4]{L. Kaestner}
\author[5]{M.W. Laschke}
\author[2]{C. Wagner}
\author[1]{S. Gekle}

\affil[1]{Biofluid Simulation and Modeling, University of Bayreuth, Bayreuth, Germany}
\affil[2]{Experimental Physics, Saarland University, Saarbrücken, Germany}
\affil[3]{Institute for Theoretical Physics, University of Innsbruck, Innsbruck, Austria}
\affil[4]{Institute for Molecular Cell Biology, Research Centre for Molecular Imaging and Screening, Center for Molecular Signaling (PZMS), Medical Faculty, Saarland University, Homburg/Saar, Germany}
\affil[5]{Institute for Clinical \& Experimental Surgery, Saarland University, Homburg/Saar, Germany}

\maketitle 

\section*{Abstract}
We investigate the margination of microparticles/platelets in blood flow through complex geometries typical for \textit{in vivo} vessel networks: a vessel confluence and a bifurcation.
Using 3D Lattice-Boltzmann simulations, we confirm that behind the confluence of two vessels a cell-free layer devoid of red blood cells develops in the channel center.
Despite its small size of roughly one micrometer, this central cell-free layer persists for up to 100~$\mu$m after the confluence.
Most importantly, we show from simulations that this layer also contains a significant amount of microparticles/platelets and validate this result by \textit{in vivo} microscopy in mouce venules.
At bifurcations, however, a similar effect does not appear and margination is largely unaffected by the geometry.
This anti-margination towards the vessel center after a confluence may explain \textit{in vivo} observations by Woldhuis \textit{et al.} [Am. J. Physiol. 262, H1217 (1992)] where platelet concentrations near the vessel wall are seen to be much higher on the arteriolar side (containing bifurcations) than on the venular side (containing confluences) of the vascular system.

\section*{Introduction}
Red blood cells (RBCs) fill up to 45 volume percent of human blood \cite{Popel2005, Pries_2008, Misbah2013, Gompper_2015, FreundRev2014} and thus represent by far the major cellular blood constituent.
Due to their high deformability, RBCs flowing through a cylindrical channel or blood vessel prefer the low-shear rate region in the center of the channel.
By hydrodynamic interactions with the red blood cells stiffer particles such as platelets, white blood cells or artificial drug delivery agents are thus expelled towards the wall.
This separation of red blood cells and stiffer particles is known as margination and is essential for the ability of blood platelets to quickly stop bleeding or for drug delivery agents to closely approach the endothelial wall.
One of the first observations of margination studied white blood cells \textit{in vivo} as well as an \textit{in vitro} model system containing disks and spheres already in 1980 \cite{SchmidSchoenbein1980}.
Interestingly, an \textit{in vivo} study by Woldhuis \textit{et al.} \cite{woldhuis1992concentration} demonstrated a striking difference between the platelet distribution in arterioles and venules, with significantly more margination occurring on the arteriolar side of the vascular system.
Since then, more detailed insights were gained by experimental studies \cite{Eckstein1988,Jain2009,Charoenphol2010, Chen2013, Namdee_2013, Wang2013, Lee_2013_margination, Fitzgibbon_2015, DApolito_2015_technique, DApolito_2015_shape, carboni_direct_2016,fay_cellular_2016}, computer simulations \cite{Migliorini2002,Freund2007, Kumar2011,  Zhao2011, tan2012influence,Zhao2012,Kumar2012, Fedosov2012,Freund_2012, Reasor2013, Kumar2014,Fedosov_Gompper_2014,  Mueller2014,Vahidkhah_2014, Vahidkhah_2015, Rivera2015, Mueller2016, Gekle2016, Krueger2016, Mehrabadi_2016, Spann_2016} as well as theoretical modeling \cite{Tokarev2011,Crowl2011, henriquez_rivera_mechanistic_2016,qi_theory_2017,qi2018time}.
These studies all deal with margination in spatially constant geometries such as shear flow, pipes with cylindrical or rectangular cross-section or plane-Couette systems.

In contrast, in living organisms blood vessels form a hierarchical structure where large arteries branch all the way down to microcapillaries in a series of bifurcations followed by a reversed series of confluences leading up to larger and larger vessels on the venular side. 
The typical distance between two bifurcations lies within 0.4~mm up to 1~mm in the microvascular system \cite{Pries_2008, Gompper_2015}. 
Despite their importance, studies on RBC distribution and margination using spatially varying geometries are surprisingly scarce.
Platelets have been studied by 2D simulations in the vicinity of an aneurysm \cite{Mountrakis2013,wu2017numerical} and in the recirculation zone behind a sudden expansion of a channel \cite{Zhao2008}.
Near a vessel constriction the locally varying distribution of RBCs \cite{Faivre_2006, Fujiwara_2009,Vahidkhah_2016_stenosis} and rigid microparticles in suspension with RBCs have been investigated \cite{Wang_2013_thrombus,Skorczewski2013, Yazdani_2016, baecher_clustering_2017}.
In ref. \cite{baecher_clustering_2017} a local increase in microparticle concentration in front of the constriction has been reported.
On a technical side, microchannels including bifurcations are investigated as a possible basis for microdevices separating blood plasma \cite{li2012blood,Tripathi2015}.
In asymmetric branches the Zweifach-Fung effect \cite{Svanes1968,Fung1973} describes an asymmetric red blood cell distribution, i.e. a larger hematocrit in the large flow rate branch \cite{pries1989red,Pries1996}. 
Under certain circumstances even an inversion of the Zweifach-Fung effect may occur \cite{Shen2016}.
Combining experiments and simulations in a rectangular channel with bifurcation \cite{Leble2011} and for a diverging and converging bifurcation using 2D simulations \cite{Wang2016} a cell-depleted zone right after the confluence has been reported.
Downstream a bifurcation an asymmetry of red blood cell distribution has been seen \cite{Leble2011,Barber_2011, li2012blood,ng2016symmetry}.
Ref. \cite{namgung2017near} reported margination of hardened RBCs while flowing through branching vessels.
\citeauthor{balogh2017direct} \cite{balogh2017direct} investigated transient behavior of red blood cell motion in more complex networks \cite{balogh2017direct,balogh_computational_2017}.
Besides the red blood cell behavior it is important to consider suspended particles like blood platelets or synthetic particles, since possible influences of bifurcations and confluences may play a major role in medical applications.
Nevertheless, systematic studies covering particle blood suspensions in networks are scare.
White blood cell motion in asymmetric bifurcations has been studied experimentally in the context of a branched vessel geometry \cite{yang2011traffic} while Sun \textit{et al.} \cite{Sun2008} investigated the interaction of six red blood cells flowing behind a white blood cell in the vicinity of vessel junctions.

In this report, we study margination of stiff spherical particles suspended among red blood cells in the vicinity of a vessel confluence and bifurcation.
This allows us first to confirm and investigate quantitatively the previously observed cell-depleted layer of red blood cells after a confluence.
Second, we provide results on the influence of network geometries on stiff particle margination.	
Our generic stiff particles are a model for artificial drug delivery agents, but also serve as a reasonable approximation for blood platelets.
We investigate two cylindrical branches either bifurcating from or forming a confluence into a larger vessel by means of 3D Lattice-Boltzmann simulations.
In order to realize simulations of these systems we implement inflow and outflow boundary conditions to the Lattice-Boltzmann/Immersed-Boundary algorithm similar to a dissipative particle dynamics based approach by Lykov \textit{et al.} \cite{lykov_inflowoutflow_2015}.
Behind a vessel confluence, we observe a RBC-free layer in the center of the channel persisting for up to 100~$\mu$m after the confluence.
Importantly, this central cell-free layer is not only devoid of red blood cells, but in addition contains a significant amount of anti-marginated microparticles/platelets.
Using fluorescent microparticles in mouse microvessels we consistently observe this anti-margination also \textit{in vivo}.
At bifurcations, no equivalent effect occurs.
Our findings may explain \textit{in vivo} observations of \citeauthor{woldhuis1992concentration} \cite{woldhuis1992concentration}, who found that in the vascular system platelet margination is strongly present at the arteriolar side with bifurcations, but less at the venular side with confluences. 
Similarly, recent observations show that thrombi formed in arterioles contain significantly more platelets than thrombi formed in venules \cite{Casa_2015} which is another indication of increased platelet margination on the arteriolar side.
By considering the axial distribution of microparticles along the flow direction we furthermore reveal the site of confluence as a spot with locally increased concentration.
The paper is organized as follows: we first introduce the simulation and experimental methods and then report two-dimensional and one-dimensional concentration profiles first in the system with vessel confluence including the experimental results then in the system with a bifurcation.
Finally, we investigate the influence of larger hematocrit as well as microparticle distribution in an asymmetric bifurcation.

\section*{Methods}

\subsection*{Lattice-Boltzmann/Immersed Boundary Method}

Fluid flow in the confluence/bifurcation geometry is modeled using a 3D Lattice-Boltzmann method (LBM) which calculates fluid behavior by a mesoscopic description \cite{Succi2001, Duenweg2008, Aidun2010}.
We use the implementation of LBM in the framework of the simulation package ESPResSo \cite{Limbach2006,Roehm2012,Arnold2013}.
Red blood cells and particles are modeled using the immersed boundary method (IBM) \cite{Peskin2002,Mittal2005,Gekle2016,baecher_clustering_2017}.

In order to mimic realistic conditions of blood flow we assign the blood plasma density $\rho_{plasma} = 1000~\text{kg}/\text{m}^3$ and viscosity $\mu_{plasma}=1.2\times10^{-3}~\text{Pas}$.
Due to the large size of red blood cells and microparticles we expect the temperature to hardly affect collective flow behavior of cells and particles and thus neglect the effect of thermal fluctuations.
A typical fluid grid of the present simulations contains 170x110x58 nodes for a bifurcation and 288x110x58 nodes for a confluence.
The time step is chosen as $0.09~\mu\text{s}$ with the time of a typical simulation being about $2.5~\text{s}$.

Red blood cells and microparticles are realized by an infinitely thin elastic membrane interacting with the fluid.
For the calculation of elastic forces imposed on the fluid, the membrane is discretized by nodes that are connected by triangles.
A red blood cell possesses 1280 triangles and 642 nodes and has a diameter of $7.82~\mu\text{m}$.
The averaged distance between neighboring nodes is about one LBM grid cell.
Nodes transfer forces to the fluid and are themselves convected with the local fluid velocity.
Interpolation between membrane nodes and fluid nodes is done using an eight-point stencil.
The viscosity contrast of the cells is $\lambda = \eta_{in} / \eta_{out} =1$, i.e. the fluid inside and outside the cells has the same viscosity.
The elastic properties of a red blood cell are achieved by applying the Skalak model \cite{BarthesBiesel2011,FreundRev2014,DaddiMoussaIder_2016_first} with a shear modulus $k_S = 5\times10^{-6}~\text{N/m}$ and an area dilatation modulus $k_A = 100 k_S$. 
Additional bending forces are computed on the basis of the Helfrich model using a bending modulus 
$k_B = 2\times10^{-19}~\text{Nm}$ \cite{FreundRev2014, Guckenberger2016,guckenberger_theory_2017}.
For the calculation the algorithm denoted method A in \cite{Guckenberger2016} is used with the bending energy being proportional to the angle of adjacent triangles and the actual forces being computed by analytically differentiating the energy with respect to node position.
This somewhat simplistic approach is appropriate for the present work where we focus on collective rather than detailed single cell behavior and where especially the behavior of the microparticles is of interest.

Microparticles are modeled in a similar fashion as the red blood cells with 320 triangles and 162 nodes.
The microparticles are chosen to have half the size of RBCs ($a=3.2~\mu\text{m}$), which has been reported to show strong margination \cite{Mueller2014}.
In contrast to the RBCs the microparticles contain an additional inner grid to ensure the stiffness and (approximate) non-deformability of the microparticles \cite{Gekle2016}.
The inner grid is linked to the membrane nodes by a harmonic potential.
Elastic properties of microparticles are chosen 1000 times larger than for RBCs.
For the purpose of numerical stability we apply an empirical volume conservation potential \cite{Krueger2016} as well as a short ranged soft-sphere repulsion, which decays with the inverse fourth power of the distance and with a cut-off radius equal one grid cell. 
The latter potential acts between all particles and between the particles and the channel wall.

Stability and accuracy of our simulation method are extensively validated in \cite{Gekle2016, Guckenberger2016} and in the Supplemental Information.
The shapes of isolated red blood cells in cylindrical and rectangular channels have been validated to agree with methodically very different Dissipative Particle Dynamics \cite{fedosov2014deformation} and Boundary-Integral \cite{GuckenbergerKihm2018} simulations.

\subsection*{Channel geometry}

\begin{figure}
	\centering
	a)
	\begin{minipage}{0.43\textwidth}
		\includegraphics[width=\textwidth]{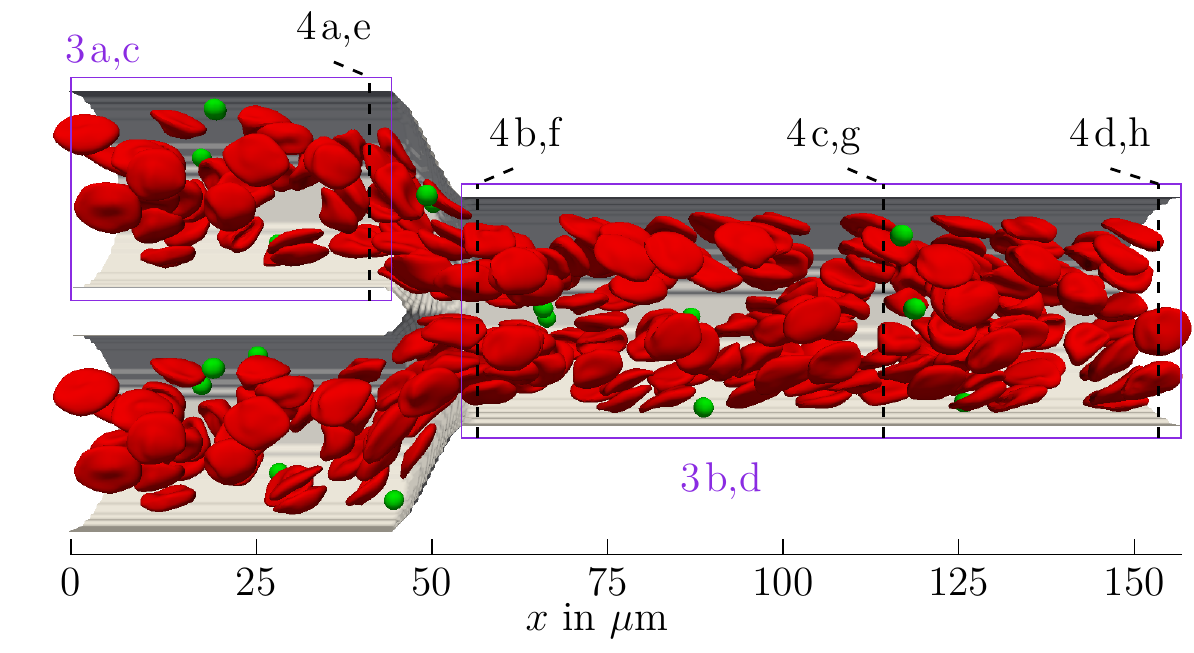}
	\end{minipage}
	\begin{minipage}{0.29\textwidth}
		\includegraphics[width=\textwidth]{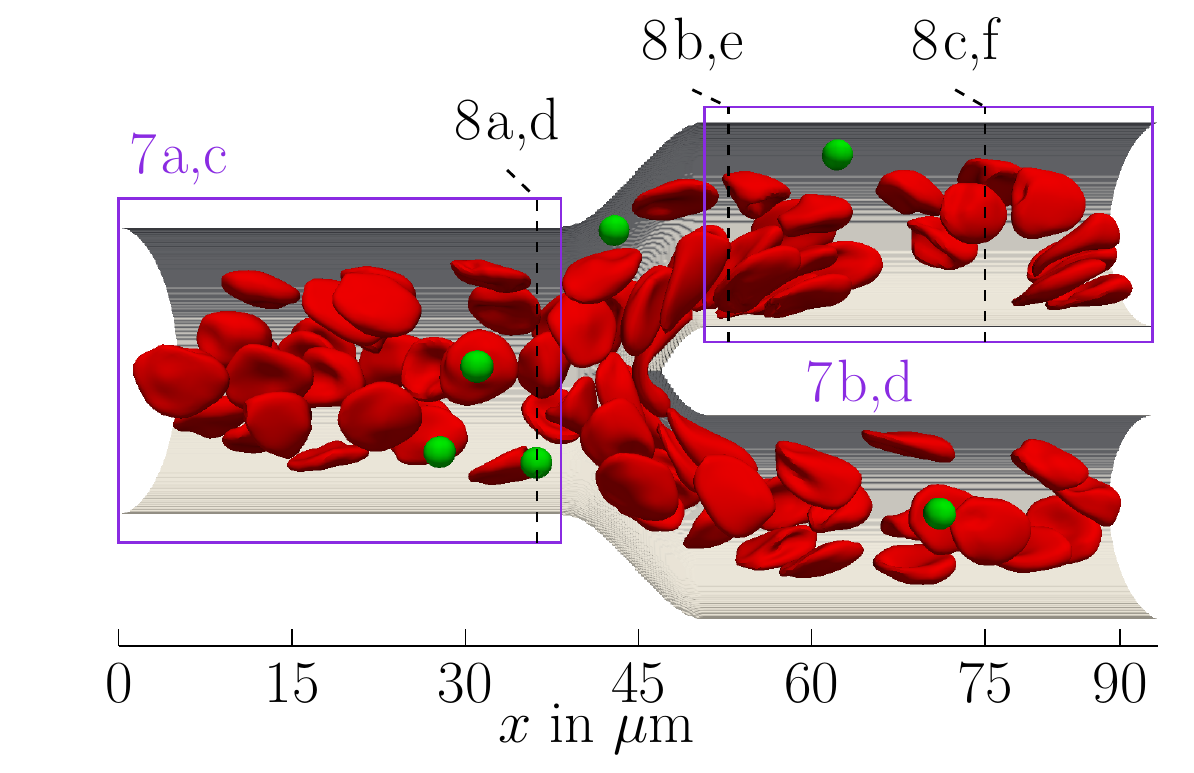}
	\end{minipage}

	\includegraphics[width=.05\textwidth]{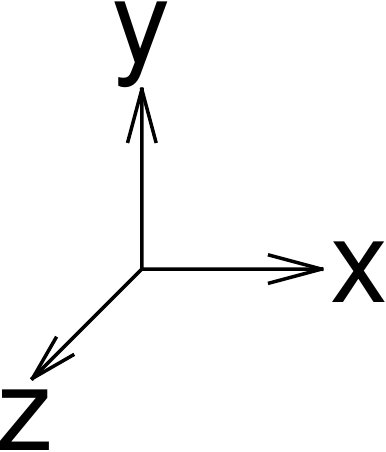}\hspace{1cm}
	b)
	\begin{minipage}{0.38\textwidth}
		\includegraphics[width=\textwidth]{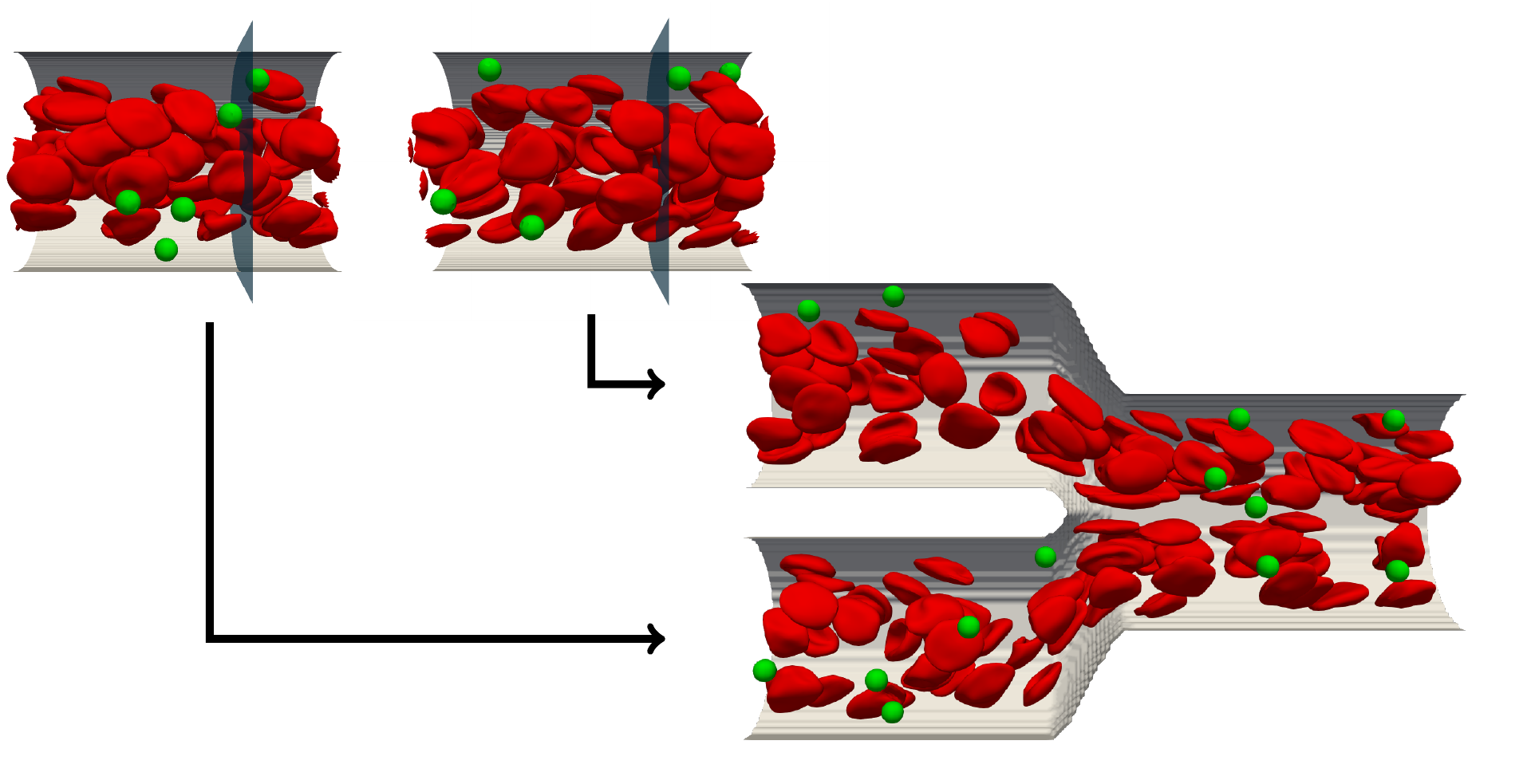}
	\end{minipage}
	\caption{
	a) Systems of interest: a suspension of red blood cells and microparticles flowing either through a confluence (left) or a bifurcation (right).
    Rectangles with numbers refer to figures containing corresponding 2D radial/planar projections while dashed lines refer to figures containing cross-sectional profiles.
	b) Inflow using straight cylinders as feeding systems: whenever a cell/particle crosses the indicated plane, it is fed into one of the branches of the confluence system.
	}
	\label{FIG:systemInflow}
\end{figure}

The systems of interest, a confluence and a bifurcation, are shown in figure \ref{FIG:systemInflow}~a).
Both geometries are constructed in the same way:
a main cylindrical channel of radius $R_{ch}$ branches into two symmetric daughter channels of radius $R_{br}$.
In order to obtain a smooth boundary, i.e., the boundary itself and the first derivative is continuous, the transition between main channel and the branches is modeled by third order polynomials:
one polynomial, $y_c (x)$, describes the bifurcating centerline, another, $y_{up}(x)$, describes the upper/lower boundary.
By rotation around the centerline with radius $y_{up}(x) - y_{c}(x)$ we obtain a circular cross-section for each $x$ forming the bifurcation along the flow direction.
Where the cross-sections of the two branches overlap the boundary is left out.  

\subsection*{Inflow/outflow boundary condition for IBM-LBM}

In order to investigate a confluence and a bifurcation as displayed in figure \ref{FIG:systemInflow}~a) periodic boundary conditions cannot be employed.
(Joining both geometries into one very large system would be computationally far too expensive due to the very long-ranged influence of bifurcations/confluences as will be detailed in the course of this work.)
At the same time, at the entrance of the branches in figure \ref{FIG:systemInflow}~a) left as well as at the entrance of the main channel in figure \ref{FIG:systemInflow}~a) right, we do not want the microparticles and red blood cells to enter in a randomly distributed fashion, but instead obey a marginated configuration in order to match with the well-known behavior in a long tube.
The purpose behind this inflow condition is to bring out clearly how the behavior of the marginated fraction of microparticles is influenced by the confluence/bifurcation.

To meet these requirements we implemented inflow and outflow boundary conditions to our IBM-LBM algorithm similar to a recent work using Dissipative Particle Dynamics \cite{lykov_inflowoutflow_2015}.
We start by simulating straight cylinders with periodic boundaries and a body force driving the flow as depicted at the top of figure \ref{FIG:systemInflow}~b).
These feeding simulations yield a time dependent sequence which then serves for particle inflow in the complex system of interest.
During the simulation of the complex system we check a frame of the feeding sequence for cells and particles crossing a certain (arbitrary) plane.
Particles crossing this plane, are then inserted at the same radial position with the same shape into the complex system as illustrated also in figure \ref{FIG:systemInflow}~b).
For crossing of the plane the center-of-mass serves as criterion similar to \cite{lykov_inflowoutflow_2015}.
The inflow velocity is chosen to match the flow rate prescribed in the straight cylinder.
In order to prevent overlap of cells during inflow we sometimes increase the flow rate slightly (about 10\%).
As a result we obtain a marginated pattern at the entrance of our complex systems as proven by the cross-sectional concentration profiles shown in the Supplemental Information.

Since azimuthal motion of the dilute microparticles in the feeding channels is extremely slow, even a very long feeding simulation would lead to a biased distribution of microparticles upon entering the complex channel. 
This is prevented by applying a small angular random force to the microparticles in the feeding channel thus guaranteeing an azimuthally homonegenous, yet well marginated distribution of microparticles.
We furthermore show in the Supplemental Information that after a first filling of the system the cell and particle number in the system reaches a plateau and slightly fluctuates around a constant value.

For the fluid, in order to prescribe a distinct flow rate we assign a constant velocity to all fluid LBM nodes at the beginning of the simulation box.
The same is done at the end of the box taking into account the different cross-sections of main channel and branches, thus matching fluid inflow and outflow.
About 15 grid cells behind the inflow the flow profile matches tube flow.
This region with evolving flow profile is skipped for particle inflow and in data analysis.

\begin{figure}[h]
	\centering
	\includegraphics{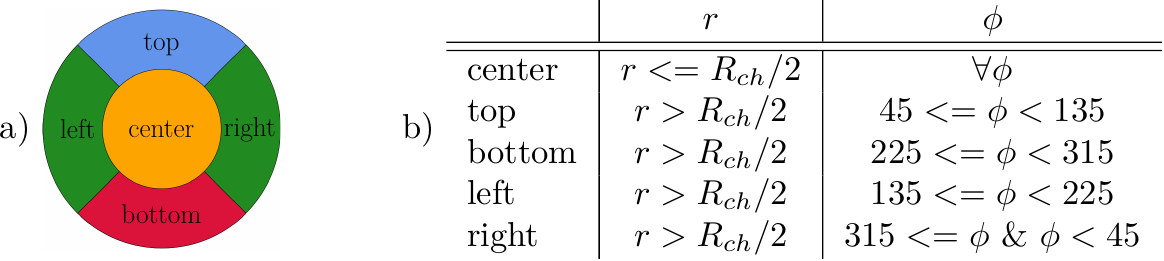}
	\caption{a) Color labeling and b) labeling criteria for red blood cells and microparticles at the entrance of the system with respect to the in-plane position in the cross-section. Due to symmetry the particles left and right can be treated equally.}
	\label{FIG:labeling}
\end{figure}

\subsection*{Analysis}

In our work, we employ different concentration profiles at given positions along the channel.
First, we compute cross-sectionally averaged concentrations leading to 1D concentration profiles as a function of position along the flow direction $x$.
Some of these profiles consider only a certain fraction of cells or particles entering the channel in specific regions which are labeled as a function of their lateral ($r$,$\phi$) position in polar coordinates at the entrance of the system.
Corresponding concentrations are calculated taking into account only this particular fraction of cells or particles.
The labeling is illustrated by the color code in figure \ref{FIG:labeling}~a) and by the criteria for $r$ and $\phi$ in figure \ref{FIG:labeling}~b).

Second, we use three types of two dimensional concentration profiles.
For microparticle concentrations, 2D radial projections in the $r, x$ plane are calculated which reflect the radial symmetry of the main and the branch channels. 
Such projections, however, are not appropriate to understand non-radially symmetric effects occuring near confluences or bifurcations.
We thus employ in addition, mainly for red blood cell concentrations, planar projections of the 3D concentrations on the $y, x$ plane by integrating the concentration over the $z$ direction perpendicular to the plane of the paper.
Finally, in order to get further insight into cell and particle distributions perpendicular to the flow direction, we calculate cross-section profiles within the $y,z$ plane.

All concentration profiles are averaged over the whole simulation time starting from the moment at which the number of cells and particles does not vary significantly.

\subsection*{Preparation of dorsal skinfold chamber and in vivo imaging}
\textbf{Animals}
The in vivo experiments were performed in 10-12 week old male C57BL/6 mice (n = 3) with a body weight of 23-26 g. The animals were bred and housed in open cages in the conventional animal husbandry of the Institute for Clinical \& Experimental Surgery (Saarland University, Germany) in a temperature-controlled environment under a 12 h/12 h light-dark cycle and had free access to drinking water and standard pellet food (Altromin, Lage, Germany). The experiment was approved by the local governmental animal care committee (approval number 06/2015) and was conducted in accordance with the German legislation on protection of animals and the NIH Guidelines for the Care and Use of Laboratory Animals (Institute of Laboratory Animal Resources, National Research Council, Washington, USA).

\textbf{Dorsal skinfold chamber model}
Microvessels were analyzed in the dorsal skinfold chamber model, which provides continuous microscopic access to the microcirculation of the striated skin muscle and the underlying subcutaneous tissue \cite{laschke2016dorsal}. For the implantation of the chamber, the mice were anesthetized by i.p. injection of ketamine (75 mg/kg body weight; Ursotamin\textsuperscript{\textregistered}; Serumwerke Bernburg, Bernburg, Germany) and xylazine (15 mg/kg body weight; Rompun\textsuperscript{\textregistered}; Bayer, Leverkusen, Germany). Subsequently, two symmetrical titanium frames (Irola Industriekomponenten GmbH \& Co. KG, Schonach, Germany) were implanted on the extended dorsal skinfold of the animals, as described previously in detail \cite{laschke2011dorsal}. Within the area of the observation window, one layer of skin was completely removed in a circular area of $\sim$15 mm in diameter. The remaining layers (striated skin muscle, subcutaneous tissue and skin) were finally covered with a removable cover glass. To exclude alterations of the microcirculation due to the surgical intervention, the mice were allowed to recover for 48 h.

\textbf{In vivo microscopy}
In vivo microscopic analysis was performed, as previously described \cite{brust2014plasma}. In detail, the mice were anesthetized and a fine polyethylene catheter (PE10, 0.28 mm internal diameter) was inserted into the carotid artery for application of the plasma marker 5 \% fluorescein isothiocyanate (FITC)-labeled dextran 150,000  (Sigma-Aldrich, Taufkirchen, Germany) and microspheres (Fluoresbrite Plain YG 1.0 $\mu$m; Polysciences, Warrington, PA, USA). Then, the animals were put in lateral decubital position on a plexiglas stage and the dorsal skinfold chamber was attached to the microscopic stage of an upright microscope (Axiotech; Zeiss, Jena, Germany) equipped with a LD EC Epiplan-Neofluar 50x/0.55 long-distance objective (Zeiss) and a 100 W mercury lamp attached to a filterset (excitation 450-490 nm, emission > 520 nm).
The microscopic images were recorded using a CMOS video camera (Prime 95B, Photometrics, Tucson, AZ, USA) at an acquisition speed of 415 images per second controlled by a PC-based acquisition software (NIS-Elements, Nikon, Tokyo, Japan).

\textbf{Trajectory analysis}
The recorded video sequence was analyzed using a single particle tracking algorithm. 
Hereby, the intensity profile of each frame was adjusted to have both the top and bottom 1\% of all
pixels saturated, correcting for changes in illumination and exposure time. 
With the aid of a tailored MATLAB script, all spherical (round) objects were detected and interconnected among all frames by cross-correlating consecutive images. 
To only detect microspheres (and not red blood cells), we set a threshold of 0.9 as a lower limit in normalized intensity values, since they are fluorescent.
Further, we defined a minimal diameter for the detected particles (0.7 µm), causing a trajectory to end if the measured value falls below this value. 
Combining the coordinates of all classified microspheres in this way over the whole video sequence,
we derived the respective trajectories.

\section*{Channel confluence}

We first investigate the system with two branches of radius $16~\mu\text{m}$ merging into one main channel of radius $17.5~\mu\text{m}$ as depicted in figure \ref{FIG:systemInflow}~a) left.
The centerlines of the two branches are separated by $39~\mu\text{m}$ and the transition zone from the end of the branches to the beginning of the main channel is about $13~\mu\text{m}$.
We choose the mean velocity at the entrance of our system to be about $v = 2.5~\text{mm/s}$.
Simulations are first performed for a physiologically realistic hematocrit (red blood cell volume fraction) of $Ht = 12\%$.
Results for a higher $Ht = 20\%$ are qualitatively similar and are presented at the end of this contribution.
The Reynolds number calculated from the centerline velocity, the red blood cell radius $R_{RBC}$, and the kinematic viscosity of the fluid $\nu_{plasma}$ is $Re = \frac{R_{RBC} \cdot v}{\nu} = \mathcal{O}(10^{-2})$.

\subsection*{Cell and particle distribution}

\begin{figure}
	\centering
	a)
	\begin{minipage}{0.35\textwidth}
		\begin{tikzpicture}
		\centering
		\node[anchor=south west,inner sep=0] (image) at (0,0) { \includegraphics[width=\textwidth]{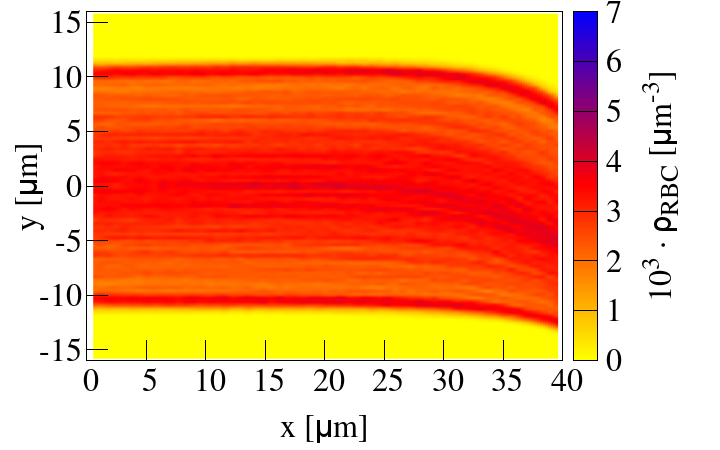} };
		\begin{scope}[x={(image.south east)},y={(image.north west)}]
		\node (RBC) at (0.7,0.93) { \scriptsize{RBC} };
		\end{scope}
		\end{tikzpicture}
	\end{minipage}
	b)
	\begin{minipage}{0.35\textwidth}
		\begin{tikzpicture}
		\centering
		\node[anchor=south west,inner sep=0] (image) at (0,0) { \includegraphics[width=\textwidth]{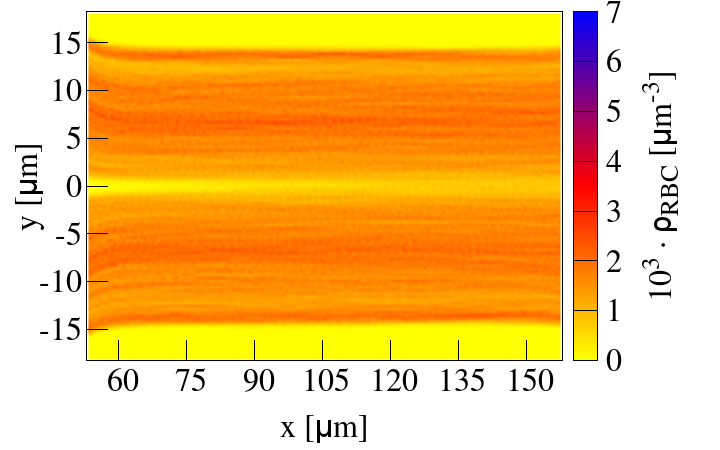} };
		\begin{scope}[x={(image.south east)},y={(image.north west)}]
		\node (RBC) at (0.7,0.93) { \scriptsize{RBC} };
		\end{scope}
		\end{tikzpicture}
	\end{minipage}
	
	c)
	\begin{minipage}{0.35\textwidth}
		\begin{tikzpicture}
		\centering
		\node[anchor=south west,inner sep=0] (image) at (0,0) { \includegraphics[width=\textwidth]{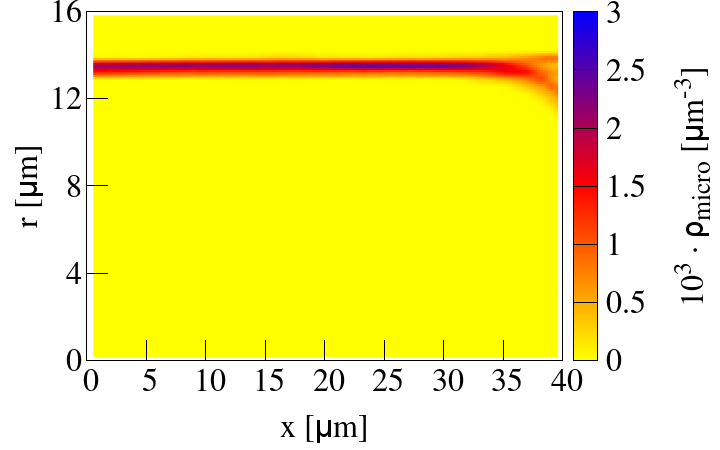} };
		\begin{scope}[x={(image.south east)},y={(image.north west)}]
		\node (micro) at (0.7,0.93) { \scriptsize{micro} };
		\end{scope}
		\end{tikzpicture}
	\end{minipage}
	d)
	\begin{minipage}{0.35\textwidth}
		\begin{tikzpicture}
		\centering
		\node[anchor=south west,inner sep=0] (image) at (0,0) { \includegraphics[width=\textwidth]{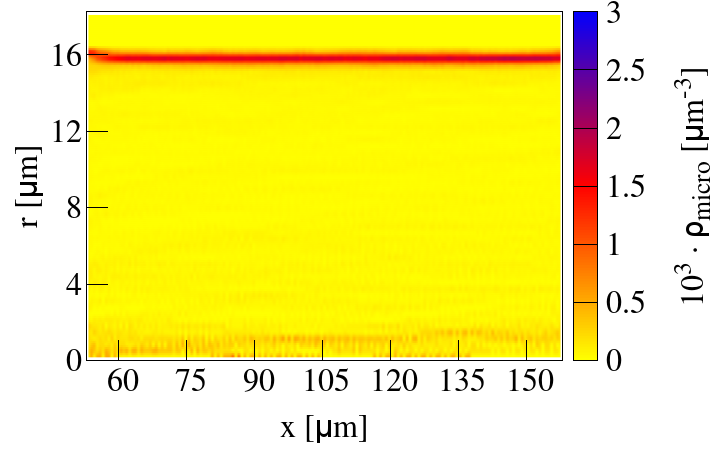} };
		\begin{scope}[x={(image.south east)},y={(image.north west)}]
		\node (micro) at (0.7,0.93) { \scriptsize{micro} };
		\end{scope}
		\end{tikzpicture}
	\end{minipage}
	\caption{
		Concentration of red blood cells in a confluence in 2D planar projection a) along the upper branch and b) along the main channel. Microparticle concentration in 2D radial projection c) along the upper branch and d) along the main channel.		
		The cell-free layer near the inner boundary decreases at the end of the branches while it increases near the outer boundary.
		Inside the main channel an additional cell-free layer in the center develops.
	}
	\label{FIG:twoDflow_merge}
\end{figure}
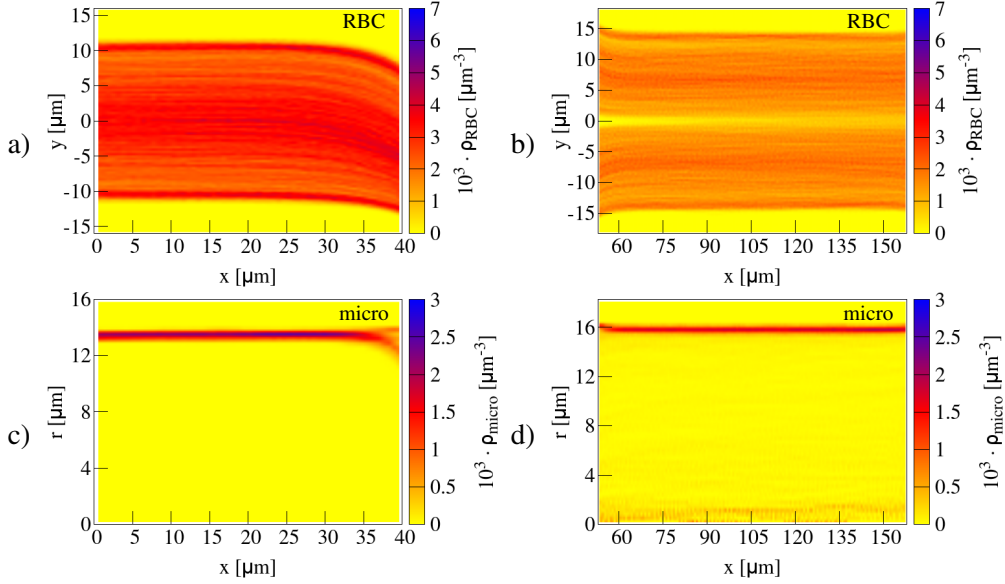

\begin{figure}[h]
	\centering
	\includegraphics[width=\textwidth]{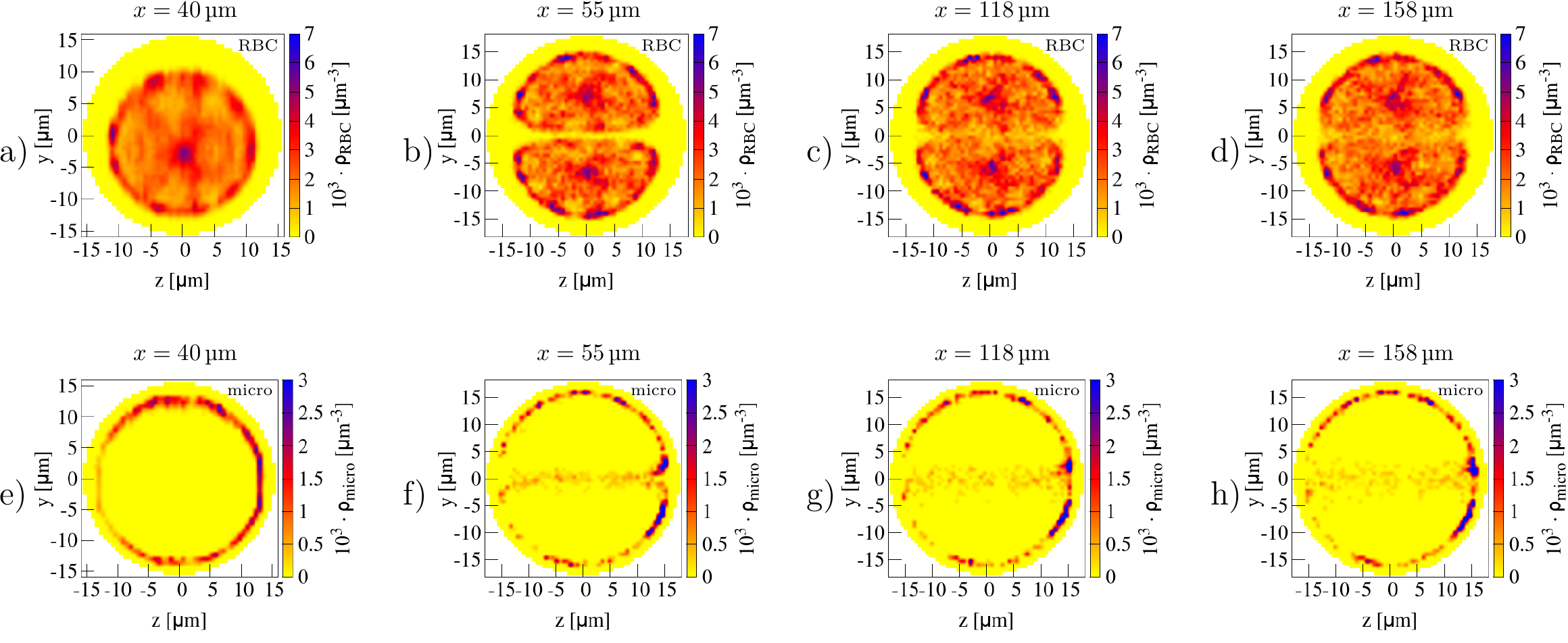}
	\caption{
		2D cross-sectional concentration in a confluence for red blood cells a)-d) and microparticles e)-h) at the end of the branches a),e), at the beginning of the main channel b),f), at the middle of the main channel c),g), and at the end of the main channel d),h). Positions are also indicated by the dashed black lines in figure~\ref{FIG:systemInflow}~(a).
		In the main channel a clear cell-free layer in the center together with microparticle anti-margination is present.
	}
	\label{FIG:twoDcross_merge}
\end{figure}

We start by considering 2D concentration profiles along flow direction in figure \ref{FIG:twoDflow_merge}.
In the two small branches and far away from the confluence we observe a homogeneous distribution of red blood cells around the center and the cell-free layer with vanishing concentration \cite{Fedosov_2010_CFL, Freund_2011, Katanov_2015} near the wall as can be seen by figure \ref{FIG:twoDflow_merge}~a).
The microparticle concentration in figure \ref{FIG:twoDflow_merge}~c) exhibits the typical margination peak near the wall.
The state of full and azimuthally homogeneous margination is confirmed by the cross-sections at channel entrance shown in the Supplemental Information.
This behavior is the same as in a straight channel.

Approaching the confluence we observe an asymmetric cell-free layer:
near the inner boundary of the branch the cell-free layer decreases, near the outer boundary it increases.
The asymmetry becomes more pronounced towards the end of the branches ($x\approx 40~\mu m$) stemming from cells flowing towards the main channel.
However, the motion of red blood cells towards the main channel already initiates at $x\approx 30~\mu m$, i.e., about $10~\mu \text{m}$ before the end of the branches.
This also affects microparticle behavior as can be seen by the two separate peaks in the radially projected concentration in figure \ref{FIG:twoDflow_merge}~c).
These two peaks stem from the particles near the inner and the outer boundary, respectively:
the microparticles near the inner boundary remain close to the wall, while the particles near the outer boundary migrate away from the wall due the flow profile towards the main channel and the increased cell-free layer.

Entering the main channel we observe a decreased cell-free layer near the upper and lower boundary in figure \ref{FIG:twoDflow_merge}~b) right at the beginning.
An interesting feature of the RBC concentration is the additional cell-free layer which develops in the channel center after the confluence.
This agrees well with the findings of ref. \cite{Leble2011} that a cell-depleted zone behind the apex of a confluence develops.
We confirm that finding by considering the concentration and in addition highlight the long-range stability of the central cell-free layer.

Most remarkably, this central cell-free layer contains, just as its classical near-wall counterpart, a significant amount of microparticles as can clearly be seen in the radially projected microparticle concentration of figure \ref{FIG:twoDflow_merge}~d).

Before investigating further this central cell-free layer, we consider cross-sectional concentration profiles in figure \ref{FIG:twoDcross_merge}.
In figure \ref{FIG:twoDcross_merge}~a), we observe how the circular pattern of red blood cells is shifted towards the inner boundary at the end of the branches.
This corresponds to the asymmetric cell-free layer in figure \ref{FIG:twoDflow_merge}~a).
Entering the main channel the pattern of red blood cell concentration possesses two flattened and asymmetric spots (figure \ref{FIG:twoDcross_merge}~b)) clearly separated by the central cell-free layer which shows vanishing concentration.
At the left and right of the main channel an additional large cell-free spot is obtained.
This central-cell-free layer stems from the cells flowing out of the upper and lower branch competing for the channel center.

Microparticles in the two branches remain well marginated until the end of the branches as shown in \ref{FIG:twoDcross_merge}~e).
After the confluence, however, figure~\ref{FIG:twoDcross_merge}~f) shows how a notable fraction of microparticles is now located very near the channel center.
This can be understood by the original location of the microparticles inside the branches:
those microparticles that are located near the inner boundary of the branch enter the main channel in the center.
This location is favorable due to the additional central cell-free layer observed in figure \ref{FIG:twoDcross_merge}~b).
In the Supplemental Information we show the anti-margination also for platelet-shaped microparticles.
Thus, the present geometry leads to a re-distribution of microparticles from a near-wall marginated position before the confluence to a near-center anti-marginated position after the confluence.

\subsection*{Lifetime of the central-cell-free layer, anti-margination, and physiological consequences}

An interesting question is the stability of the RBC-depleted central cell-free layer and the corresponding anti-marginated microparticles as the flow continues away from the confluence location along the main channel.
Figure \ref{FIG:twoDflow_merge}~b) shows that the central-cell-free layer is surprisingly stable, being visible all along the main channel and only becoming slightly blurred towards the end.
The same trend can be observed in the cross-sectional concentration at three sites along the main channel in figure \ref{FIG:twoDcross_merge}~b),c),d).
Although it starts to become blurred after 60~$\mu$m in figure \ref{FIG:twoDcross_merge}~c), the central cell-free layer and especially the cell-free spot left and right is visibly present until at least 100~$\mu$m behind the confluence as shown in figure~\ref{FIG:twoDcross_merge}~d).
Similarly, the corresponding microparticle concentration in figure \ref{FIG:twoDcross_merge}~f),g),h) shows that microparticles are located in the center all along the main channel.
Thus, a confluence of two channels influences microparticle behavior over distances which are much longer than the channel diameter.

To gain a more mechanistic insight into this long-time stability of the central CFL, we calculate the (shear-induced) diffusion coefficient \cite{turitto1975platelet,goldsmith1979flow,Zhao2011,Kumar2012Margination,Zhao2012,Grandchamp_2013,Katanov_2015} of red blood cells in the center.
For this, we compute the time-dependent mean squared displacement (MSD) which is shown in the Supplemental Information.
By modeling the increase in MSD with time by the theoretical expectation for normal diffusion $\braket{ \Delta y(t)^2 } = 2Dt$ 
we extract a diffusion coefficient for the red blood cells of $D_{RBC} \approx 28~\mu\text{m}^2 / \text{s}$ in the case of $Ht$ = 12\%.
This value is of the same order as previous results in experiments with red blood cells \cite{turitto1975platelet,goldsmith1979flow,Grandchamp_2013} and simulations of spheres and platelets \cite{Zhao2011,Zhao2012}.
By assuming a thickness of $1.5~\mu$m and a flow speed of 2.5~mm/s we calculate a distance of 100~$\mu$m required to bridge the central cell-free layer.
This length scale agrees well with the observation in the concentration profiles that the central cell-free layer starts to become blurred after 100~$\mu$m.
We calculate in the same way the shear-induced diffusion coefficient of the microparticles $D_{micro} \approx 25~\mu\text{m}^2 / \text{s}$.
Ref. \cite{turitto1975platelet} reports for platelets in a perfusion chamber $34~\mu$m$^2$/s for a shear rate of $832~$1/s and unknown hematocrit.
Ref. \cite{Zhao2011} and \cite{Zhao2012} obtain a diffusivity a factor 2 smaller in simulations of plane-Couette flow with $Ht = 0.2$.
Considering not only the gap of the central cell-free layer to be closed, but the larger spot left/right (assuming a distance of $5~\mu$m to be bridged) we can estimate a distance of 1.1 mm for red blood cell redistribution, which is comparable to the estimation of $25D$ by \citeauthor{Katanov_2015} \cite{Katanov_2015}.	
When we estimate the length scale required to migrate towards the channel wall, i.e. to marginate, we get about 5~mm.
Comparing this to the typical distance between successive confluences of about 0.4-1~mm \cite{Pries_2008, Gompper_2015} we conclude that full margination cannot be regained. This in turn may explain \textit{in vivo} observations that on the venular side of the vascular system margination is much less pronounced than on the arterioral side \cite{woldhuis1992concentration, Casa_2015}.

Furthermore, we want to address the question how strong the effect of anti-margination is.
Therefore, we calculate the fraction of particles that are not located directly next to the wall, i.e. we consider particles that are more than one particle radius away from the vessel wall. 	
For the concentration profiles in fig. \ref{FIG:twoDcross_merge} we obtain the fractions 0 at $x=40~\mu$m, 0.158 at $x=55~\mu$m, 0.138 at $x=118~\mu$m, and 0.135 at $x=158~\mu$m.
We find that the fraction decreases very slowly with increasing distance from the confluence because of marginating microparticles (most likely those located left/right).

\begin{figure}[h]
	\centering
	\includegraphics[width=.45\textwidth]{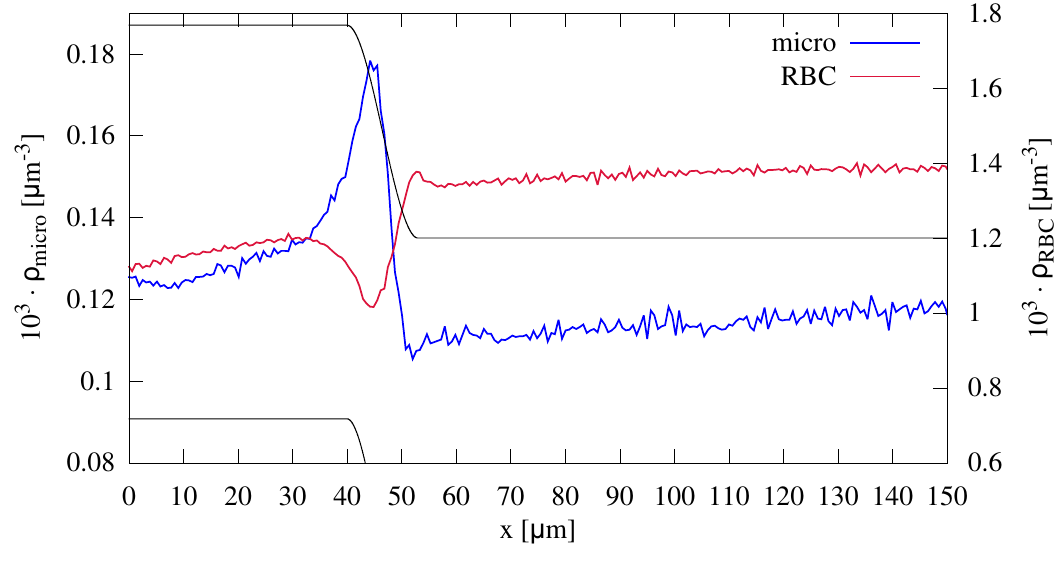}
	
	\caption{1D axial profile of red blood cells and microparticles flowing through a vessel confluence.
	While the red blood cells are depleted at the site of the confluence, the microparticles exhibit a concentration increase of about 50\% compared to the branches and the main channel.	
	}
	\label{FIG:axial_merge}
\end{figure}

\subsection*{Axial concentration}

Further insight can be gained by considering 1D axial cell and particle concentration profiles in figure \ref{FIG:axial_merge}.
The overall red blood cell concentration exhibits two plateaus inside the main channel and inside the branches, respectively.
Inside the branches the concentration is lower than in the main channel in agreement with the Fahraeus effect \cite{secomb_blood_2017}.
Right at the confluence, we observe a zone where the RBCs become slightly depleted.
The microparticle concentration along the branches first increases slightly and right at the confluence a strong peak develops.
In the main channel the microparticles have a nearly constant concentration. 

\begin{figure}[h]
	\centering
	a)
	\begin{minipage}{0.4\textwidth}
		\includegraphics[width=\textwidth]{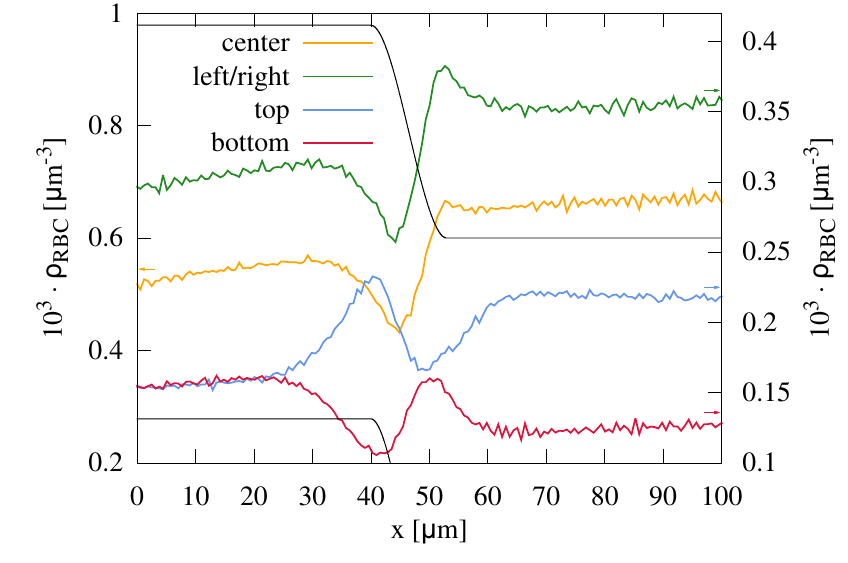}
	\end{minipage}
	b)
	\begin{minipage}{0.4\textwidth}
		\includegraphics[width=\textwidth]{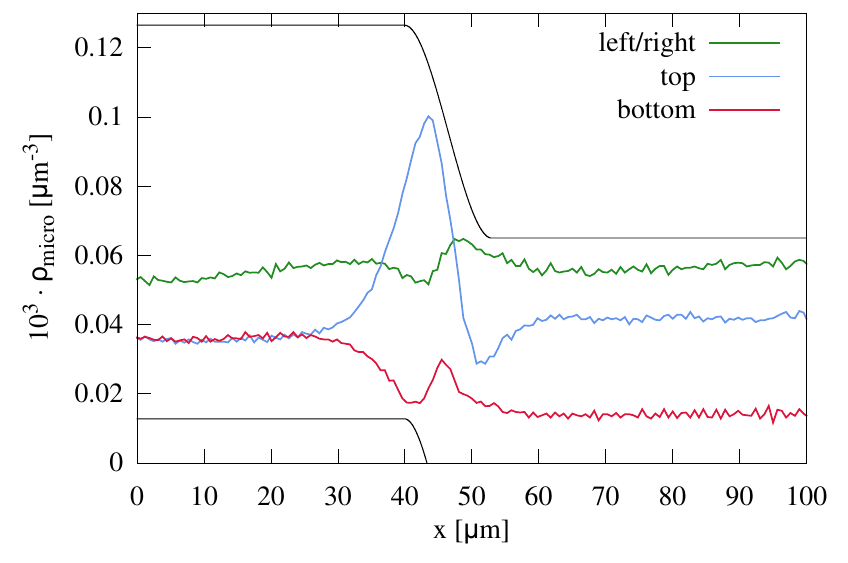}
	\end{minipage}
	
	\caption{Axial concentration of a) red blood cells and b) microparticles distinguished regarding their position inside the cross-section of the branches as illustrated in figure~\ref{FIG:labeling}.
		The microparticles entering at the top of the upper branch (or, equivalently, the bottom of the lower branch) exhibit a pronounced peak.
	}
	\label{FIG:axialLabel_merge}
\end{figure}

In order to elucidate red blood cell behavior further and, especially, in order to explain the peak in microparticle concentration we label the cells/particles while entering the branches as explained in the Methods section.
The concentrations for the labeled cells and particles are shown in figure \ref{FIG:axialLabel_merge}.

The red blood cells in the center, those left/right and those at the bottom of the branches behave in a similar fashion:
at the end of the branches they are accelerated and thus have a decreased concentration.
After a small peak, they quickly reach a constant concentration in the main channel.
Only the cells arriving at the top exhibit a slightly increasing concentration at the end of the branches, but are depleted, as well, at the site of the confluence.
Microparticles arriving left/right or at the bottom show a similar concentration profile as the corresponding red blood cells and thus do not cause the peak in overall concentration in figure \ref{FIG:axial_merge}.
We note that these concentration profiles can be understood by passive tracer particles similar as in constricted channels \cite{baecher_clustering_2017} and as detailed in the Supplemental Information.
From figure \ref{FIG:axialLabel_merge}~b) we are thus able to conclude that the peak stems from the microparticles flowing at the top of the branches.
The concentration of these microparticles increases more than two-fold compared to the branches and the main channel.
Due to the margination the microparticles are located right besides the wall. 
Also the concentration profile of the microparticles at the top can be reproduced by passive tracer particles as done in the Supplement Information.
Thus, the local increase in microparticle concentration can be understood by the underlying flow profile.


\subsection*{In vivo observation of microparticle anti-margination}
\begin{figure}[h]
	\centering
	\includegraphics[width=.6\textwidth]{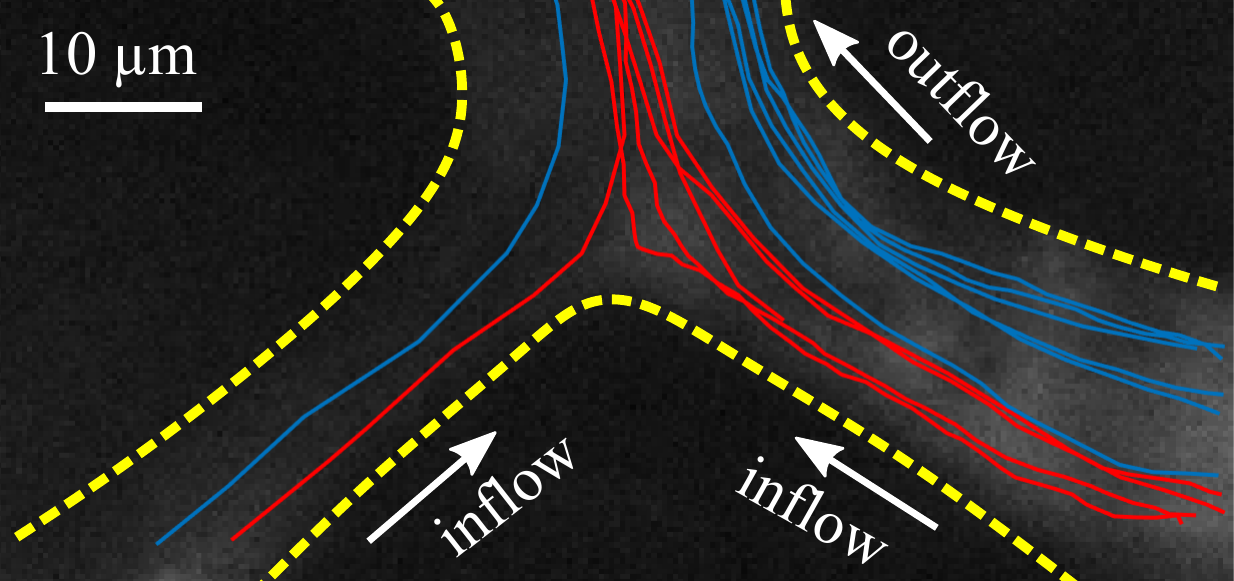}
	\caption{ \textit{In vivo} measurement of tracked fluorescent beads in mouce microvessels. 
			Blue lines show trajectories of beads which remain marginated after the bifurcation while red lines show beads undergoing anti-margination.
			Yellow dashed lines denote the vessel boundaries.
	}
	\label{FIG:experiment}
\end{figure}

To demonstrate the relevance of the anti-margination observed in our simulations, we inject fluorescent beads into living mice and image their behavior when flowing through a microvessel confluence.
In figure~\ref{FIG:experiment} we show a set of trajectories obtained from the video microscopy images (a corresponding movie is included as Supplemental Information).
In agreement with the predictions of our numerical simulations, beads which are initially marginated at the outer walls (blue lines in figure~\ref{FIG:experiment}) remain marginated whereas beads located initially at the inner walls (red lines in figure~\ref{FIG:experiment}) undergo anti-margination and end up near the channel center after passing through the confluence.

\section*{Channel bifurcation}

Next, a bifurcation is investigated as depicted in figure \ref{FIG:systemInflow}~a) right.
The suspension of red blood cells and microparticles flows through a straight channel of radius $16~\mu\text{m}$ branching into two daughter channels of radius $11.5~\mu\text{m}$.
Main channel and the combined branches have the same cross-sectional area and the centerlines of the two branches are separated by $34~\mu\text{m}$.
The transition zone from the end of the main channel to the beginning of the branches is about $13~\mu\text{m}$.

\begin{figure}[h]
	\centering
	a)
	\begin{minipage}{0.35\textwidth}
		\begin{tikzpicture}
		\centering
		\node[anchor=south west,inner sep=0] (image) at (0,0) { \includegraphics[width=\textwidth]{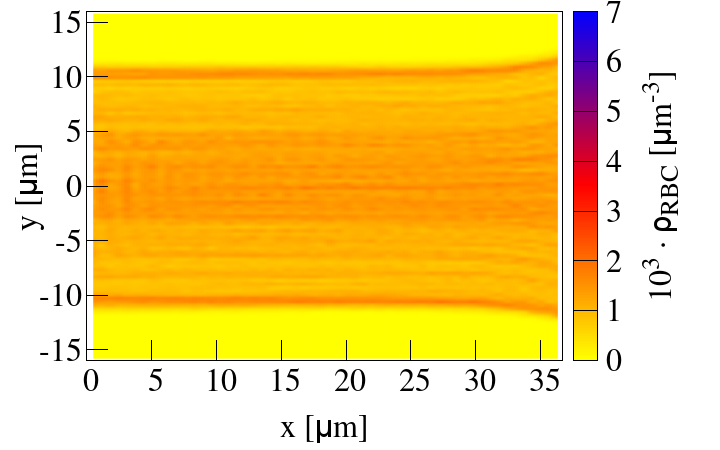} };
		\begin{scope}[x={(image.south east)},y={(image.north west)}]
		\node (RBC) at (0.7,0.93) { \scriptsize{RBC} };
		\end{scope}
		\end{tikzpicture}
	\end{minipage}
	b)
	\begin{minipage}{0.35\textwidth}
		\begin{tikzpicture}
		\centering
		\node[anchor=south west,inner sep=0] (image) at (0,0) { \includegraphics[width=\textwidth]{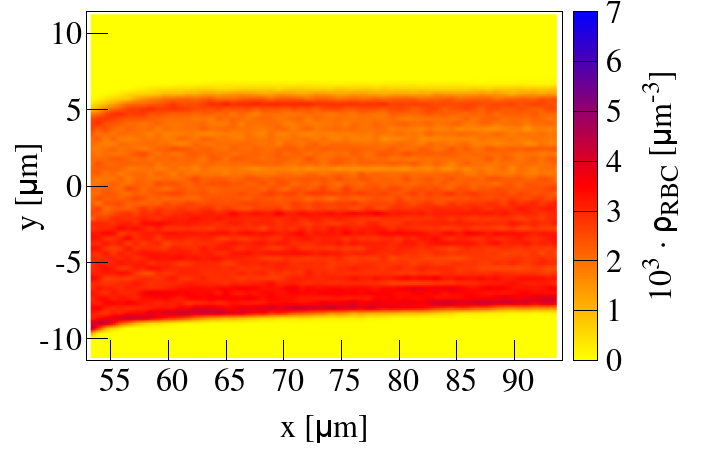} };
		\begin{scope}[x={(image.south east)},y={(image.north west)}]
		\node (RBC) at (0.7,0.93) { \scriptsize{RBC} };
		\end{scope}
		\end{tikzpicture}
	\end{minipage}
	
	c)
	\begin{minipage}{0.35\textwidth}
		\begin{tikzpicture}
		\centering
		\node[anchor=south west,inner sep=0] (image) at (0,0) { \includegraphics[width=\textwidth]{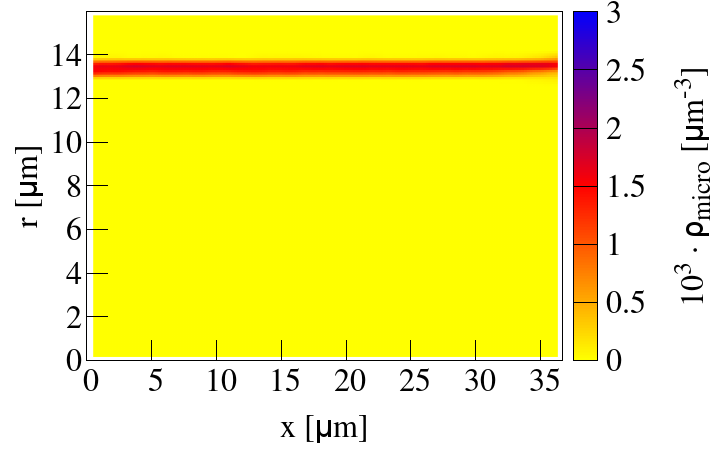} };
		\begin{scope}[x={(image.south east)},y={(image.north west)}]
		\node (micro) at (0.7,0.93) { \scriptsize{micro} };
		\end{scope}
		\end{tikzpicture}
	\end{minipage}
	d)
	\begin{minipage}{0.35\textwidth}
		\begin{tikzpicture}
		\centering
		\node[anchor=south west,inner sep=0] (image) at (0,0) { \includegraphics[width=\textwidth]{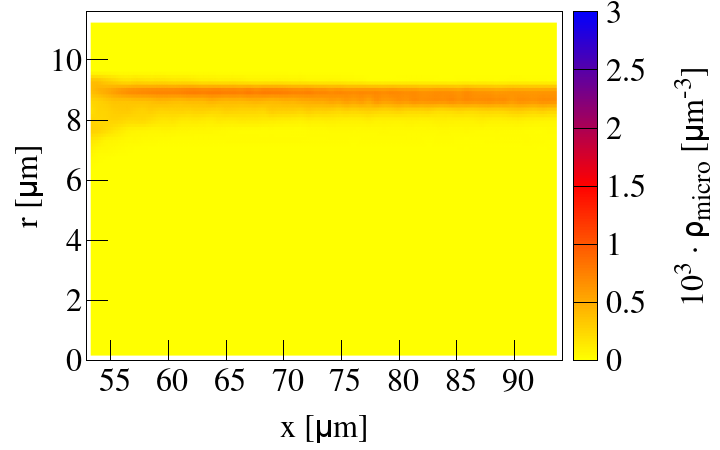} };
		\begin{scope}[x={(image.south east)},y={(image.north west)}]
		\node (micro) at (0.7,0.93) { \scriptsize{micro} };
		\end{scope}
		\end{tikzpicture}
	\end{minipage}

	\caption{
		2D planar projection within the bifurcation for red blood cells a) along the main channel and b) along the upper branch. 2D radial projection of microparticle concentration c) along the main channel and d) along the branches.
		The cell-free layer decreases at the end of the main channel and we observe an asymmetric cell distribution inside the branches.
		The margination peak of microparticles is somewhat blurred after the bifurcation, but otherwise unaffected.
	}
	\label{FIG:twoDflow_bif}
\end{figure}
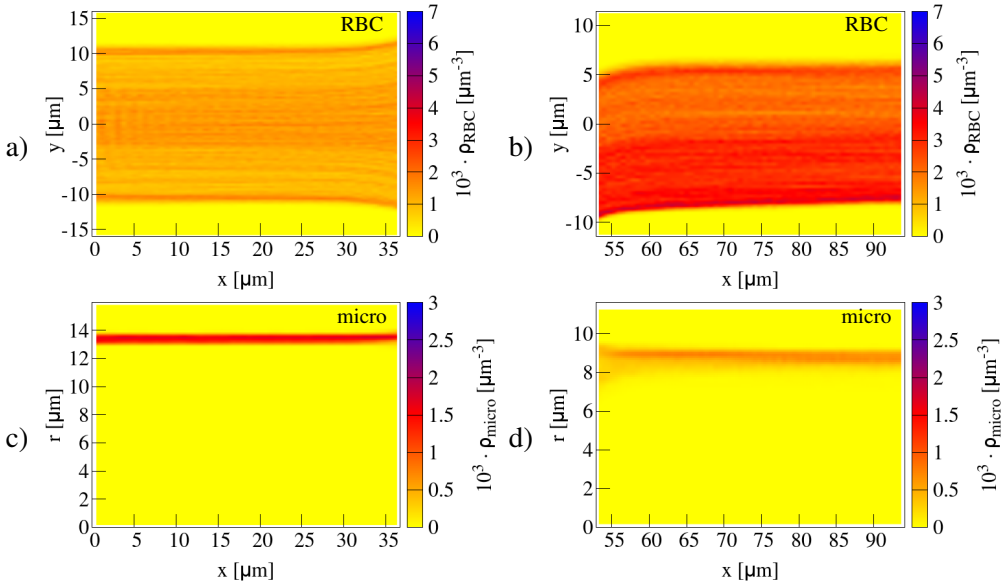

\begin{figure}[h]
	\centering
	\includegraphics[width=0.75\textwidth]{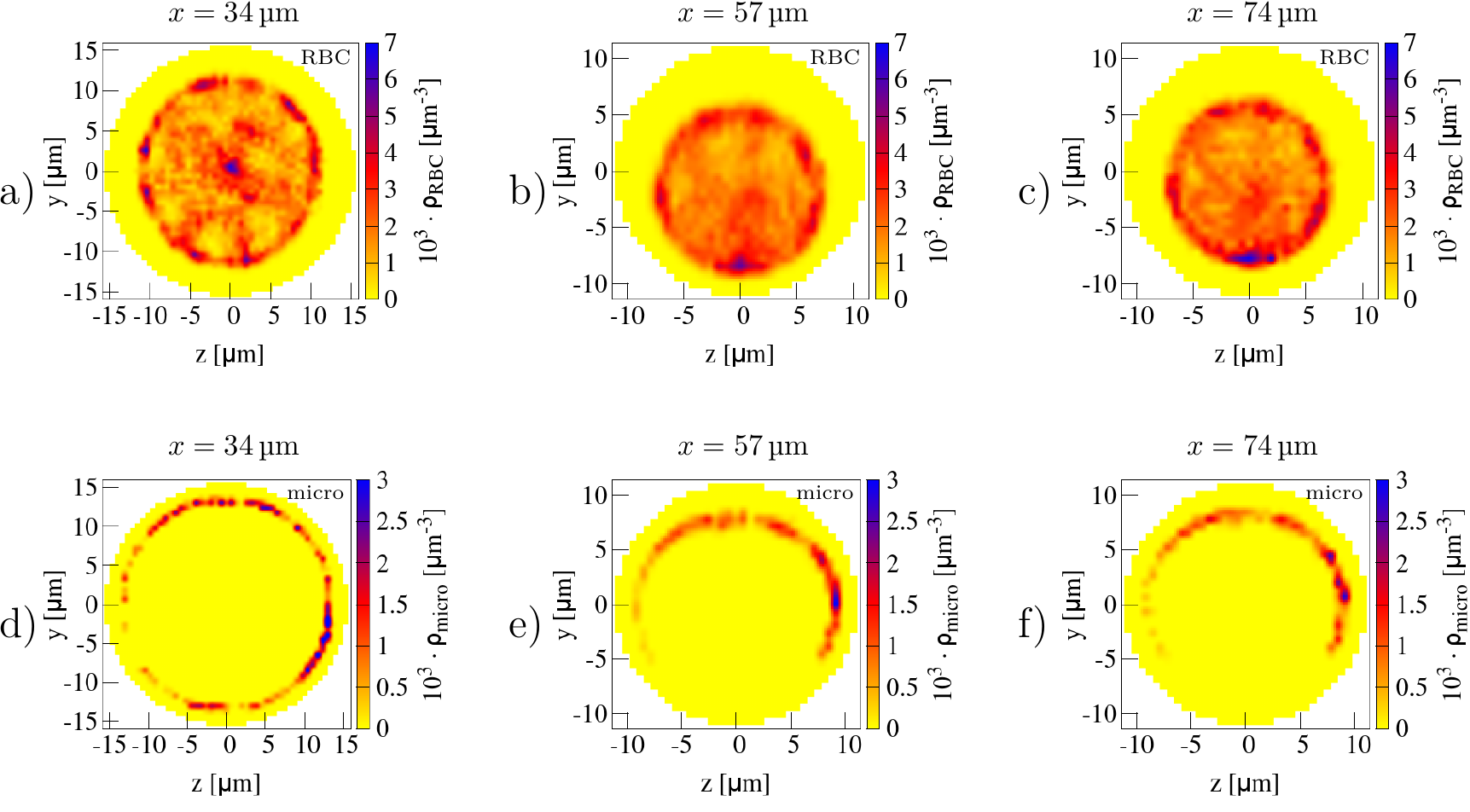}
	\caption{
		2D cross-sectional concentration of red blood cells a),b),c) and microparticles d),e),f) at the end of the main channel a),d), at the beginning of the branches b),e), and at the end of the branches c),f). See figure~\ref{FIG:systemInflow}~(a) for indications of the respective positions along the channel.
		Inside the branches an asymmetric cell-free layer develops and microparticles suffer a loss of concentration directly besides the inner wall.
	}
	\label{FIG:twoDcross_bif}
\end{figure}

\subsection*{Cell and particle distribution}

In figure \ref{FIG:twoDflow_bif} we first investigate the two-dimensional concentration of red blood cells and microparticles along the flow direction.
At the very beginning around $x\approx 0$ we again observe a homogeneous red blood cell distribution (figure \ref{FIG:twoDflow_bif}~a)) around the center and the cell-free layer with vanishing red blood cell concentration near the wall.
Approaching the bifurcation the cell-free layer decreases.
The decrease in cell-free layer is of the same amount at both locations near the upper boundary and near the lower boundary. 
It can be straightforwardly explained by the bifurcating geometry which causes the red blood cells to flow upwards/downwards into the daughter channels.
This motion into the daughter channels starts already about 10 $\mu$m before the end of the main channel and makes the cells migrate towards the outer wall. 

An asymmetry in the cell-free layer occurs inside the daughter channels (figure \ref{FIG:twoDflow_bif}~b)) as also observed in recent work \cite{li2012blood,ng2016symmetry}.
The asymmetry is especially pronounced at the beginning of the branches stemming from cells flowing in the center of the main channel, which enter either the upper or lower branch near the inner wall.
We observe a strongly decreased cell free layer near the inner boundary right at the beginning of the daughter branches.
The thickness of the cell-free layer near the outer boundary increases correspondingly.
After about 10~$\mu$m the inner and outer cell-free layers both reach a constant value, which is similar to the length scale for the re-establishment of the outer cell-free layer after the confluence.
Interestingly, the inner and outer cell-free layers do remain asymmetric and retain this asymmetry until the very end of our channel about 50~$\mu$m behind the bifurcation.

Although the cell-free layer decreases at the end of the main channel hardly an effect is observed on microparticle behavior.
Entering the branches the microparticle concentration peak only becomes more blurred because of the microparticles located near the upper boundary entering the larger cell-free layer inside the branch.

The asymmetries in cell and particle distribution can be seen in more detail in the 2D cross-section profiles in figure \ref{FIG:twoDcross_bif}.
At the end of the main channel in figure \ref{FIG:twoDcross_bif}~a) the red blood cell distribution is still circular with only small deviations corresponding to the decrease in cell-free layer seen in figure \ref{FIG:twoDflow_bif}~a).
At the beginning of the branch the circular red blood cell concentration is strongly shifted towards the inner boundary in figure \ref{FIG:twoDcross_bif}~b) and a less pronounced, but still clearly visible asymmetry is still present at the end of the branch in figure \ref{FIG:twoDcross_bif}~c).
Furthermore, a local spot with increased red blood cell concentration is observed near the inner boundary.

While the microparticles are hardly affected at the end of the main channel (figure \ref{FIG:twoDcross_bif}~d)) a notable effect is the vanishing concentration of microparticles near the inner boundary of the branches in figure \ref{FIG:twoDcross_bif}~e) and f).
Over an angle-range of about 90 degrees at the bottom of figure \ref{FIG:twoDcross_bif}~e) and f) the microparticle concentration vanishes completely.
The vanishing microparticle concentration can be understood by the radial distribution in the main channel due to margination:
in order to reach positions near the lower boundary of the branch the microparticles would have to be located near the center of the main channel, which is not the case because of margination.
Thus, we report a region within the branches that posses vanishing microparticle concentration in comparison to a simple straight channel.

\begin{figure}[h]
	\centering
	\includegraphics[width=.45\textwidth]{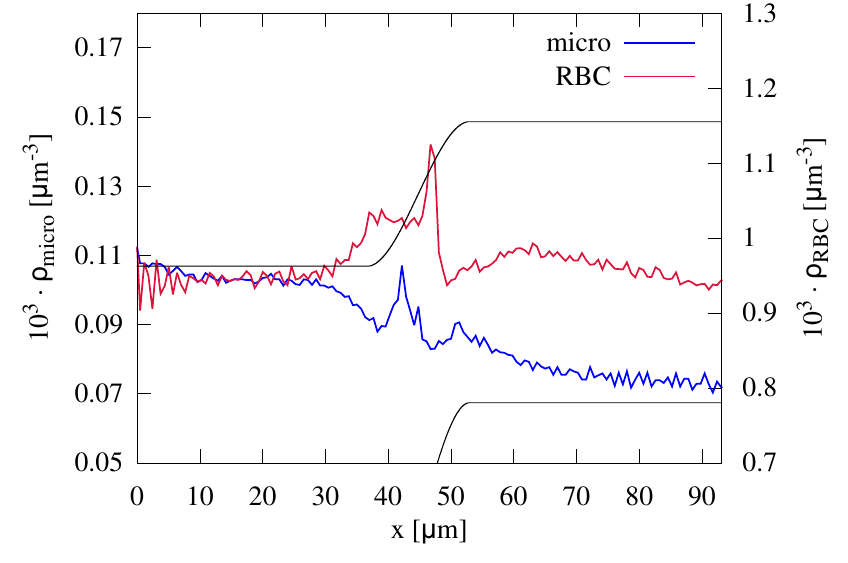}
	
	\caption{Axial concentration of red blood cells and microparticles flowing through a bifurcating channel.
		Both red blood cells and microparticles exhibit a peak in front of the bifurcation apex.
		The microparticle concentration increases directly in front of the apex, while the red blood cell concentration exhibits a second small peak due to a second cell flowing onto a cell already being stuck at the apex.}
	\label{FIG:axial_bif}	
\end{figure}

\subsection*{Axial concentration}

As for the confluence, we now investigate the behavior of cells and particles along the varying geometry by 1D axial concentration profiles in figure~\ref{FIG:axial_bif}.
After a constant plateau inside the main channel both red blood cells and microparticles show a clear peak ahead of the apex of the bifurcation.
Inside the branches the red blood cells take the same concentration as in the main channel while the microparticle concentration decreases.
The latter can be explained by the intrinsic velocity profile:
flowing besides the boundary the stiff microparticles cover a certain ring of tube diameter along the boundary.
Due to the fixed particle size this ring has the same diameter in the main channel and within the branches.
Assuming a Poiseuille flow and averaging over such a ring around the boundary leads to a higher flow rate inside the branches, since they cover a larger part of the steep velocity profile.
Thus, within the branch the microparticles experience a larger velocity, leading to lower residence time thereby causing a decreasing concentration. 

\begin{figure}[h]
	\centering
	a)
	\begin{minipage}{0.4\textwidth}
		\includegraphics[width=\textwidth]{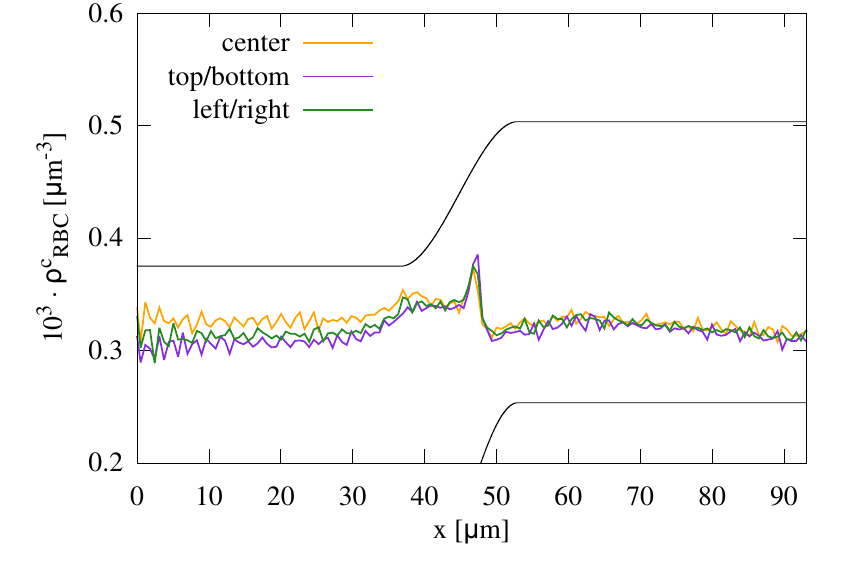}
	\end{minipage}
	b)
	\begin{minipage}{0.4\textwidth}
		\includegraphics[width=\textwidth]{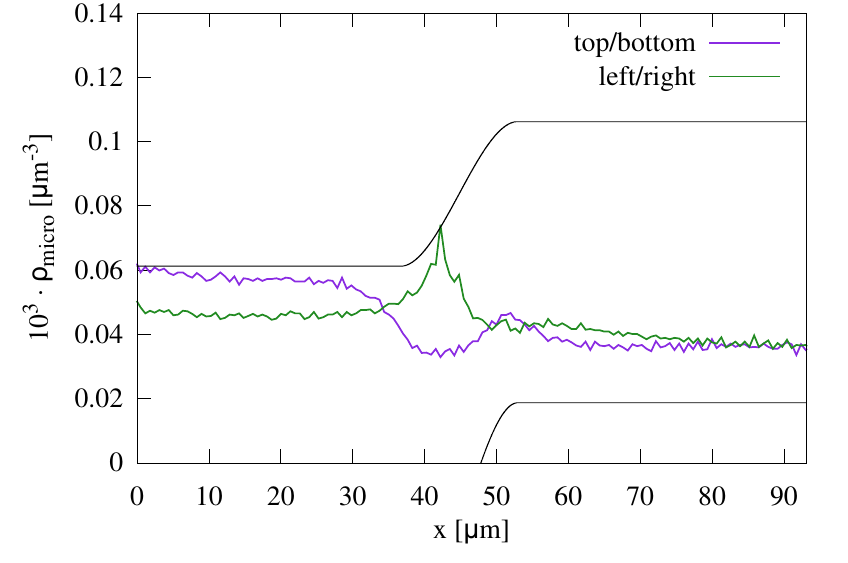}
	\end{minipage}

	\caption{Axial concentration of a) red blood cells and b) microparticles distinguished regarding their initial position inside the cross-section of the main channel.
		The peak in red blood cell concentration stems from the cells trapped at the apex of the bifurcation.
		The microparticles exhibit a similar peak when arriving left/right.
		However, for microparticles entering top/bottom no peak occurs.
	}
	\label{FIG:bif_label}
\end{figure}

We furthermore distinguish the cells and particles regarding their position inside the main channel and calculate the axial concentration profiles for each cell/particle fraction in figure \ref{FIG:bif_label}~a) and b).
All fractions of red blood cells behave in a similar manner with a small peak at the apex.
This peak stems from cells being trapped at the apex of the bifurcation: arriving in the center of the main channel, red blood cells have to break symmetry and decide for one branch.
As visible in figure \ref{FIG:systemInflow}~a) right some cells flow directly onto the apex and are trapped there before flowing in one of both branches, a phenomenon called ''lingering'' in \cite{balogh2017direct}.
We note that also the red blood cells at the top/bottom show a peak, because they are still located close enough to the center to be influenced by more central cells getting trapped at the apex.

In full analogy, a similar peak is observed for microparticles located left/right in figure \ref{FIG:systemInflow}~b).
Also these microparticles flow onto the apex and become trapped for a short period of time.
In contrast, the microparticles located top/bottom are diluted at the bifurcation.
After a subsequent little dip in concentration, microparticles from both regions quickly reach a constant concentration inside the branches.

The concentration profile of both cells and particles can be understood again by considering tracer particles (see Supplemental Information).

\section*{Influence of hematocrit}

In the following we present simulations which have the same geometrical properties as the channels in figure \ref{FIG:systemInflow} but with a hematocrit of $Ht = 20 \%$ for the inflow.

\begin{figure}
	\centering
	\includegraphics[width=\textwidth]{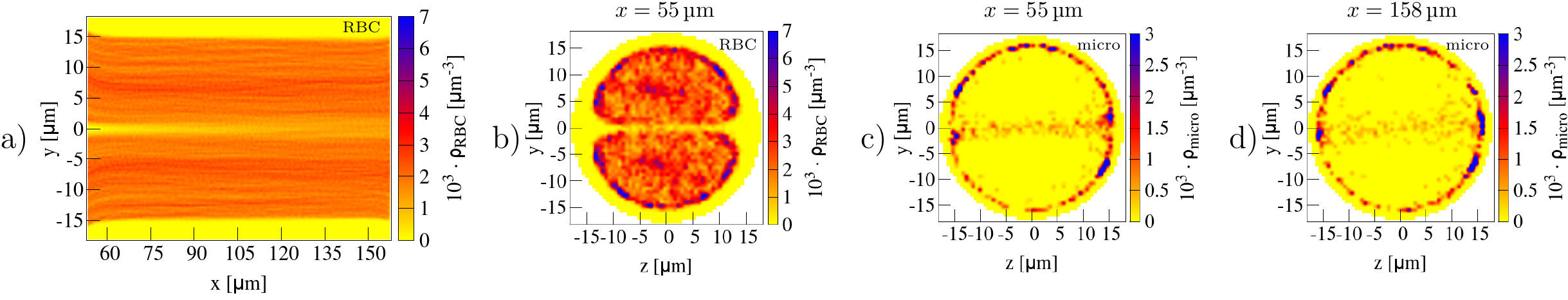}
	\caption{a) 2D planar projection within the confluence for red blood cells along the main channel behind a confluence as in figure \ref{FIG:twoDflow_merge}, but with larger hematocrit $Ht=20\%$.
b)-d)	2D cross-section profiles for b) red blood cells and c),d) microparticles in the main channel as in figure \ref{FIG:twoDcross_merge}, but for larger hematocrit $Ht = 20\%$.}
	\label{FIG:HttwoD:confluence}
\end{figure}

\begin{figure}[h]
	\centering
	a)
	\begin{minipage}{0.4\textwidth}
		\includegraphics[width=\textwidth]{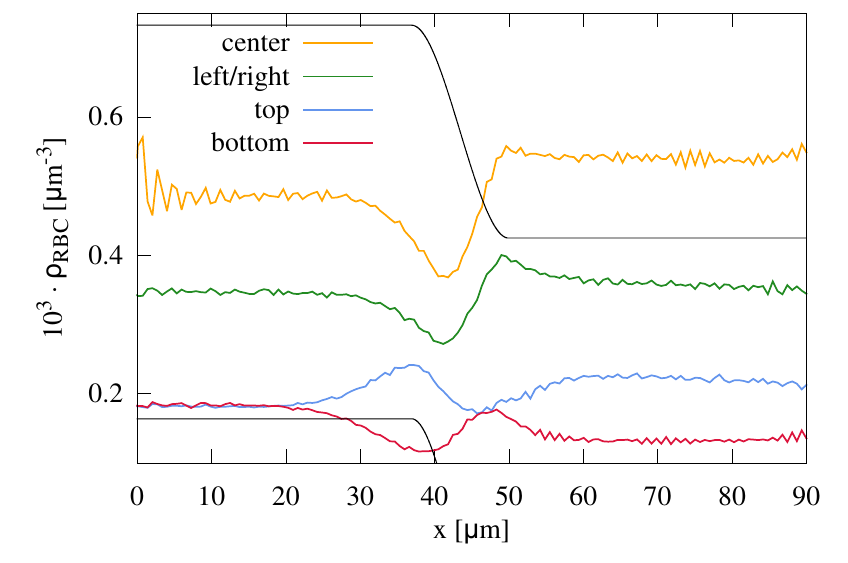}
	\end{minipage}
	b)
	\begin{minipage}{0.4\textwidth}
		\includegraphics[width=\textwidth]{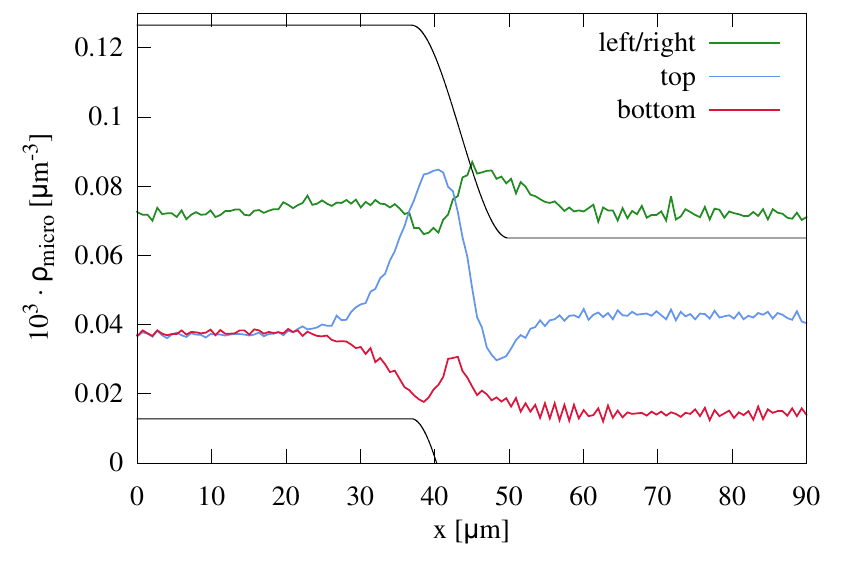}
	\end{minipage}

	c)
	\begin{minipage}{0.4\textwidth}
		\includegraphics[width=\textwidth]{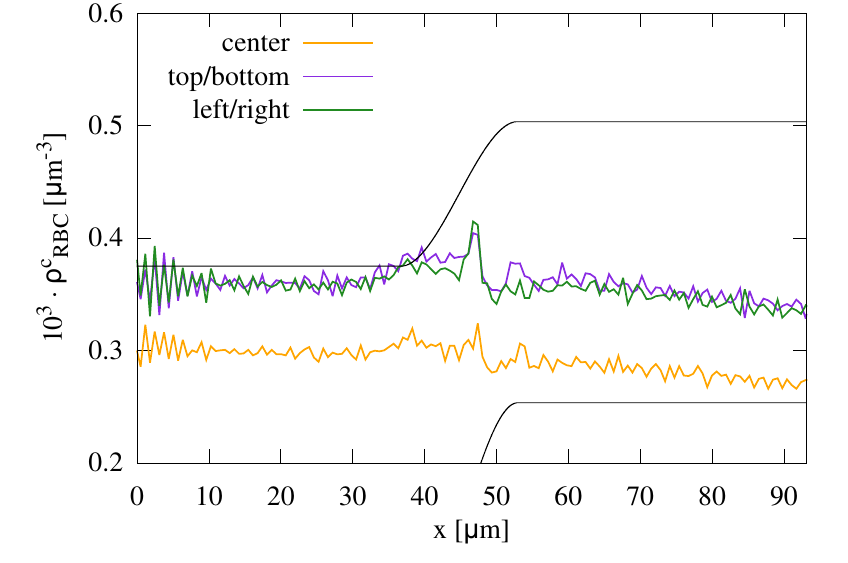}
	\end{minipage}
	d)
	\begin{minipage}{0.4\textwidth}
		\includegraphics[width=\textwidth]{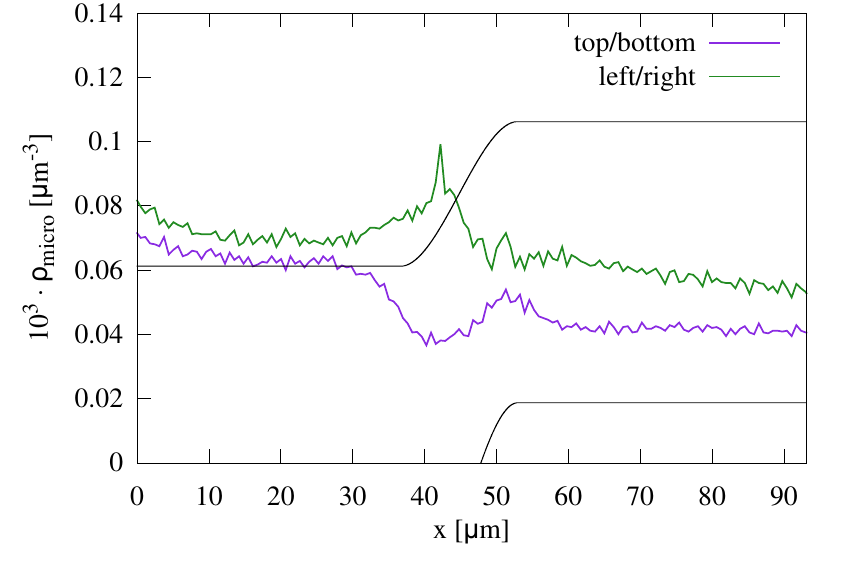}
	\end{minipage}
	\caption{ 1D axial concentration of a),c) red blood cells and b),d) and microparticles flowing through a a),b) confluence or c),d) bifurcation with larger hematocrit $Ht = 20 \%$ labeled by their radial position at the entrance.
	The larger hematocrit hardly affects the behavior of red blood cells and microparticles.		
	}
	\label{FIG:HtLabel}
\end{figure}

Figure \ref{FIG:HttwoD:confluence} shows that behind the confluence of two branches the red blood cell distribution behaves qualitatively similar as for the lower hematocrit.
Although the cell-free layer near the vessel wall is reduced compared to figure \ref{FIG:twoDflow_merge}~b) the cell-free layer in the center of the main channel in figure \ref{FIG:HttwoD:confluence}~a) is of about the same size.
Only at the left and right of the cross-section the cell-free space clearly reduces compared to lower hematocrit (figure \ref{FIG:HttwoD:confluence}~b)).
The central cell-free layer is very pronounced up to 40~$\mu$m behind the confluence, but becomes blurred slightly faster towards the end of the channel when compared to the low hematocrit case.
This faster decay can be explained by the larger shear-induced diffusion coefficient of $D_{RBC} = 38~\mu$m$^2$/s.
This agrees with the theoretical expectation that the shear-induced diffusion coefficient depends on the number of cell-cell collision and thus on the cell concentration \cite{da_cunha_hinch_1996,Grandchamp_2013}.
The microparticles are still located in the cell-free layer in the center all along the main channel as can be seen in figure \ref{FIG:HttwoD:confluence}~c) and d). 
We can again calculate the fraction of anti-marginated microparticles and obtain the fractions 0.145 at $x=55~\mu$m and 0.137 at $x=158~\mu$m.
Also when we compare the axial concentration of labeled red blood cells and microparticles in figure \ref{FIG:HtLabel}~a) and b) to the case of lower hematocrit in figure \ref{FIG:axialLabel_merge} we see similar behavior in both cases.

When we investigate the influence of larger hematocrit on the system with bifurcation, we find that each cell-free layer in the system decreases with increasing hematocrit (results shown in the Supplemental Information). 
At the end of the main channel the cell-free layer still decreases and the pronounced asymmetry of cell-free layers within the branches is present. 
The increase in concentration due to the apex of the bifurcation remains unaffected by larger hematocrit as seen in figure \ref{FIG:HtLabel}~c) and d). 
Since a certain number of cells or microparticles stacks at the apex of the bifurcation the effect is not modified when more cells are added to the system.
Especially, the absolute number of cells in the center region stays approximately the same for larger hematocrit.

\section*{Asymmetric bifurcations}

We finally touch briefly on the subject of asymmetric bifurcations.
For this, we keep the main channel radius $R_{ch} = 16~\mu\text{m}$ and the upper branch $R_{br} = 11.5~\mu\text{m}$ as in figure \ref{FIG:systemInflow} right and only vary the diameter of the lower branch.
Two different simulations are done with radius $R^{low} = 8~\mu\text{m}$ and $5.5~\mu\text{m}$.
Here, we focus on the total concentration of red blood cells and microparticles within the two branches as listed in table \ref{TAB:asymBranches}.
\begin{table}[!h]
	\centering
	\begin{tabular}{l|c|c|c|c}
		$R^{low}$ & $\rho_{RBC}^{up}$ & $\rho_{RBC}^{low}$ & $\rho_{micro}^{up}$ & $\rho_{micro}^{low}$ \\
		\hline\hline
		$8~\mu\text{m}$ & 1.027 & 0.794 & 0.0711 & 0.0792 \\
		$5.5~\mu\text{m}$ & 1.005 & 0.674 & 0.0722 & 0.0806
	\end{tabular}
	\caption{Concentration of red blood cells and microparticles in the upper (up) and lower (low) channel of an asymmetric bifurcation with radius $11 \mu\text{m}$ of the upper branch and $R^{low}$ of the lower branch. Concentrations are given in $10^3 \mu\text{m}^{-3}$.}
	\label{TAB:asymBranches}
\end{table}

The red blood cell concentration clearly differs between the upper and lower branch. The upper branch, being the one with larger flow rate, receives clearly more cells than the lower branch.
This effect is enhanced when the diameter and correspondingly the flow rate further decreases as in the case of $R^{low} = 5.5~\mu\text{m}$.
The fraction of concentration between lower and upper branch $\rho^{low}/\rho^{up}$ changes from $0.77$ to $0.67$ when changing $R^{low}$ from 8 to 5.5~$\mu$m.
We note that the total flow rate at the outflow of the system is the same in both simulations to match the flow rate at the entrance.
As a consequence, the flow rate in the upper branch slightly differs in both simulations (the fraction of flow rates is 0.5 and 0.26, respectively).
The asymmetric distribution of red blood cells qualitatively matches with the Zweifach-Fung effect observed earlier \cite{pries1989red,lykov_inflowoutflow_2015,secomb_blood_2017,balogh_computational_2017} and can be attributed to the cell-free layer \cite{secomb_blood_2017} combined with RBC deformability.

In contrast to the asymmetric red blood cell distribution microparticles are nearly evenly distributed to the daughter channels.
Furthermore, the distribution is not affected by decreasing the diameter of the lower branch, the fraction in both cases being about $1.11$.
Since the stiff microparticles are located within the cell-free layer the different flow rate does not affect their distribution.
Although the lower branch is significantly smaller the apex of the bifurcation and thus the separation line between the two branches is located near the center of the main channel by construction of the geometry.
Thus, arriving near the wall only the microparticles located around the equator are drawn into the upper branch by the flow while all other microparticles may distribute equally into the branches.
All in all, the microparticles exhibit a very similar concentration in both branches.

\section*{Conclusion}

We used a 3D Lattice-Boltzmann-Immersed-Boundary method with inflow/outflow boundary conditions to investigate a mixed suspension of red blood cells and stiff particles flowing through a vessel confluence as well as a vessel bifurcation.
The stiff particles can be regarded as models for synthetic drug delivery agents or naturally occuring stiff cells such as platelets.
In agreement with earlier studies, we observe and quantify the formation of a pronounced central cell-free layer behind the confluence of two vessels.
We find that the central cell-free layer is very stable being still observable even 100~$\mu$m after the confluence.
As a consequence, we show that stiff particles at the confluence are strongly redistributed. 
Although all stiff particles arrive on a marginated position inside the well-known near-wall cell free layer, while transversing the confluence a significant fraction of them undergoes anti-margination ending up trapped in the central cell-free layer near the channel center.
This position is retained even longer than the 100~$\mu$m life time of the central cell-free layer itself.
Calculating the fraction of anti-marginated microparticles we found that more than 13\% of the particles are located around the center 100~$\mu$m behind a confluence.	
Under the assumption that at the succeeding confluence this fraction of microparticles is still anti-marginated we estimate that after 5 confluences half of the initially completely marginated particles is now evenly distributed across the cross-section of the channel.
In contrast, a bifurcating geometry is found to not significantly influence the margination propensity of stiff particles.
For the confluence, we also conducted \textit{in vivo} measurements which proved the relevance of anti-margination of stiff microparticles in living mice.

In previous \textit{in vivo} studies platelets have been observed to be mainly located near the wall in arterioles \cite{tangelder1985distribution,woldhuis1992concentration} but not in venules where the platelet concentration was rather continuous across vessel diameter.
In a similar direction, the recent work of Casa \textit{et al.} \cite{Casa_2015} found that thrombi were platelet rich on the arterial but not on the venous side of the blood vessel network.
Our present findings may provide an explanation for these observations.
On the arterial side, the microvascular network consists mainly of bifurcations from larger into smaller and smaller vessels which, according to our findings, do not significantly disturb the margination propensity of platelets.
On the venous side, however, small channels frequently merge into larger ones. 
At such confluences, our results clearly demonstrated anti-margination, i.e., a tendency of platelets to be forced into the center of the vessel.
In a network with a cascade of confluences being only 400-1000~$\mu$m apart \cite{Pries_2008, Gompper_2015} the platelet margination near the channel wall will be further and further disturbed, ending up finally in a rather continuous concentration profile and thus explaining the experimental observations of ref. \cite{woldhuis1992concentration} and \cite{Casa_2015}.

\section*{Author Contributions}

C.B. performed research, analyzed data, wrote the manuscript.
L.S. contributed to simulation tools.
S.G. designed research, wrote the manuscript.
A.K. performed image treatment and data analysis, C.W. designed and interpreted research.
L.K. and M.W.L. designed and performed the \textit{in vivo} experiments.

\section*{Acknowledgments}

The authors thank the Gauss Centre for Supercomputing e.V. for providing
computing time on the GCS Supercomputer SuperMUC at the Leibniz
Supercomputing Centre.

\noindent This work was supported by the Volkswagen Foundation.

\noindent We gratefully acknowledge the Elite Study Program Biological Physics and the \textit{Studienstiftung des deutschen Volkes}.  

\section*{Supplemental Information}

An online Supplemental Information to this article can be found by visiting BJ Online at http://www.biophysj.org.

\bibliographystyle{abbrvnato}
\bibliography{./paper.bib}

\newpage

\listoffigures

\newpage

\listoftables



\section*{Supplementary material}

# List of Figures



# List of Tables

# Supplemental Information - Anti-margination of microparticles and platelets in the vicinity of branching vessels

C. Bächer, A. Kihm, L. Schrack, L. Kaestner, M.W. Laschke, C. Wagner, and S. Gekle

## Inflow: full margination and constant particle number

In figure S1 we show the state of complete margination at the entrance of a bifurcating channel and a confluence, respectively. The cross-sectional concentrations show the red blood cells accumulated around the channel center and the microparticles close to the wall.

Figure S2 shows that after a short transient time the implemented particle inflow/outflow leads to a constant number of red blood cells and microparticles in the system.

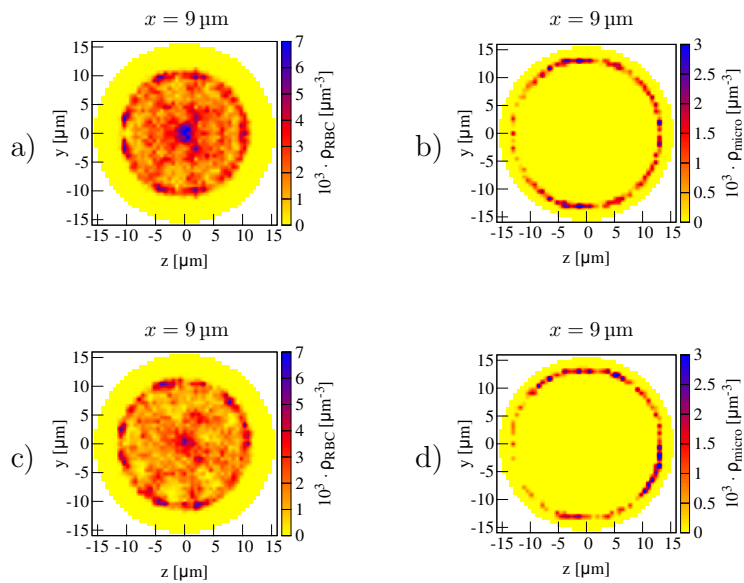

Fig. S1: At the entrance of the confluence system a),b) and the bifurcation system c),d) we have a state of full margination: the red blood cells a),c) are located in the channel center, the microparticles b),d) near the wall.

## Concentration profiles of tracer particles

In order to model the behavior of red blood cells and microparticles we use passive point particles as tracers flowing with the intrinsic velocity profile [1]. We first start with the confluence. Homogeneously distributed particles exhibit a similar concentration profile as the red blood cells. As a consequence, red blood cell behavior can be explained by the intrinsic velocity profile. Doing the same calculations for tracers in the different regions resemble the labeled red blood cells. It also fits the concentration profiles for the microparticles.

In the diverging bifurcation starting at top the concentration profile of the point particles matches that of the microparticles quite well. Also the point particles located right reproduce microparticle behavior. We note that

1

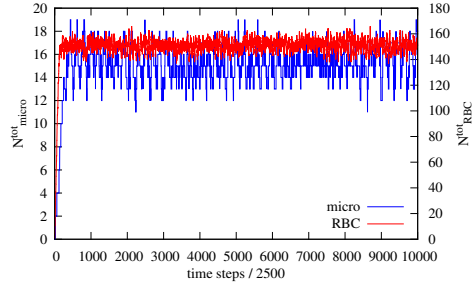

Fig. S2: Total number of RBCs and microparticles within the confluence system depending on the simulation time. Both numbers fluctuate around a constant value after initial filling of the system.

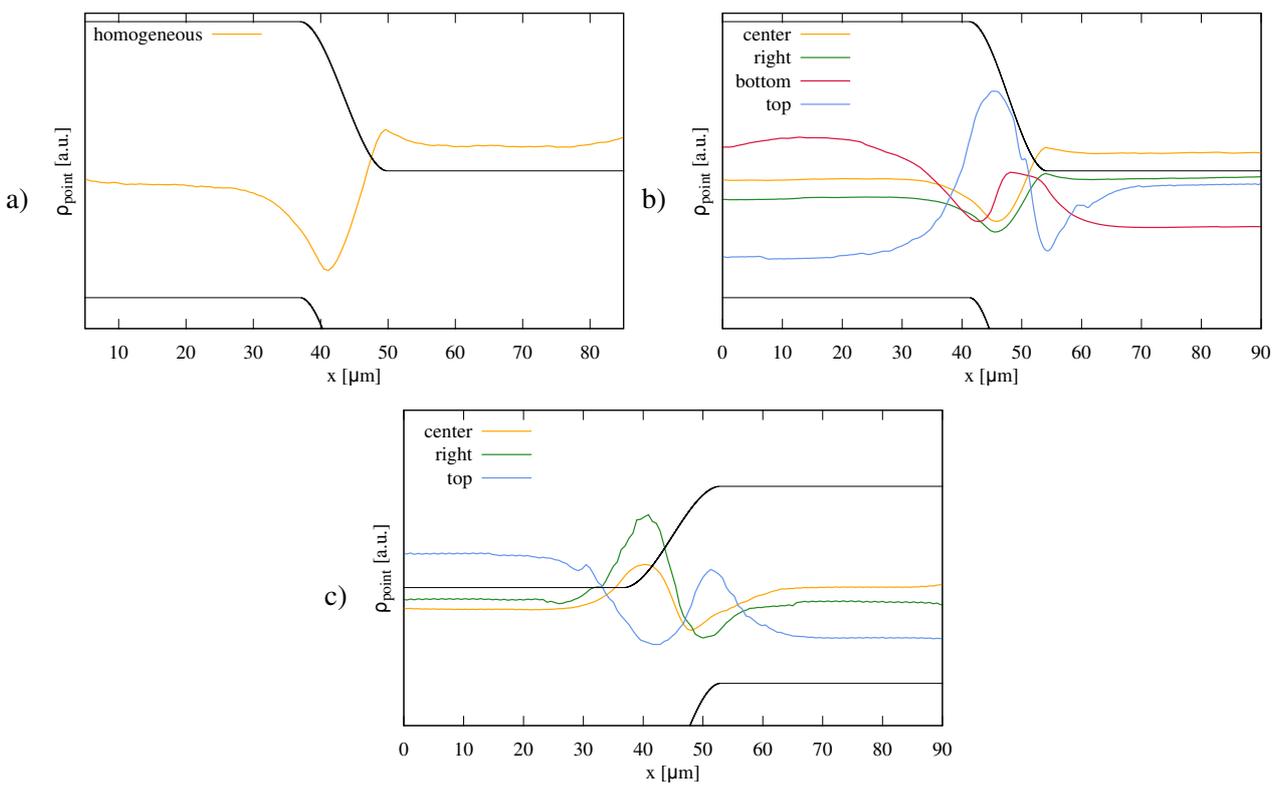

Fig. S3: Concentration of a) homogeneously distributed tracer particles and b) tracer particles flowing in the distinct regions at system entrance within the confluence. c) Tracer particles flowing in the distinct regions at system entrance within the diverging bifurcation. These figures are compared to cell and particle concentration in figure 6, 11 and 13 in the main text.

starting point particles top or bottom and left or right results in the same concentration due to symmetry. Red blood cell behavior is also similar to that of point particles, except the peak at the bifurcation apex. The differences are effects due to the finite size and deformability of red blood cells.

2

# Shear-induced diffusion

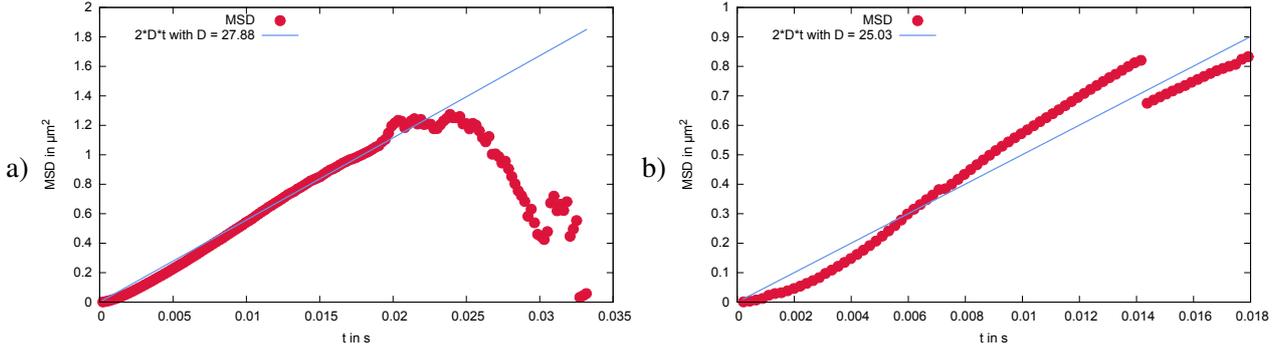

Fig. S4: Mean square displacement over time for a) red blood cells and b) microparticles located near the center behind a confluence. By modeling the theoretical expectation we can extract a shear-induced diffusion coefficient.

# Larger hematocrit

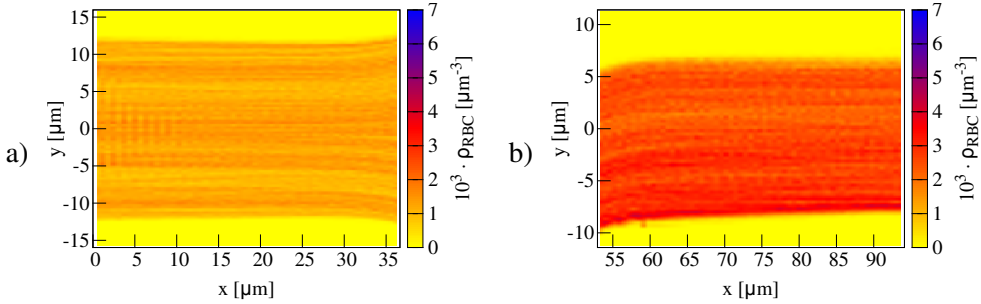

Fig. S5: 2D planar projection for red blood cells along the a) main channel and b) branches of the bifurcation with larger hematocrit $Ht = 20\%$. The cell-free layer decreases, but the behavior is qualitatively unchanged.

# Anti-margination of platelet-shaped microparticles

In the main text we focus on spherical microparticles. Here, we show additional results for oblate spheroids – a geometry that mimicks more closely that of real platelets. The platelets have a diameter of 3.9 $\mu$m along the two long axes and 2.3 $\mu$m along the small axis and are illustrated in figure S6. Similar to the spherical particles of the main text, about 14% of the spheroidal microparticles are anti-marginated directly behind the confluence.

# Validation of the used IBM-LBM algorithm

In the following, we summarize and extend the validation for our immersed boundary method (IBM) and Lattice-Boltzmann method (LBM). In ref. [3] the calculation of shear and bending forces has been validated for a capsule in shear flow. In ref. [2] the hematocrit profile for tube flow and plane-Poiseuille flow has been shown to agree with previous, established studies. Furthermore, the stability of the stiff spherical particles used has been demonstrated and the flow profile past a sphere has been compared favorably to the analytical solution.

In addition, we here calculate here the Stokes drag $1/(6\pi\eta a)$ that relates the force on a sphere of radius $a$ to its velocity in a suspending fluid of viscosity $\eta$ for a sphere with two different particle resolutions in figure S7. We performed simulations with the resolution used in the main text (81 nodes of the inner stiff grid) and an

3

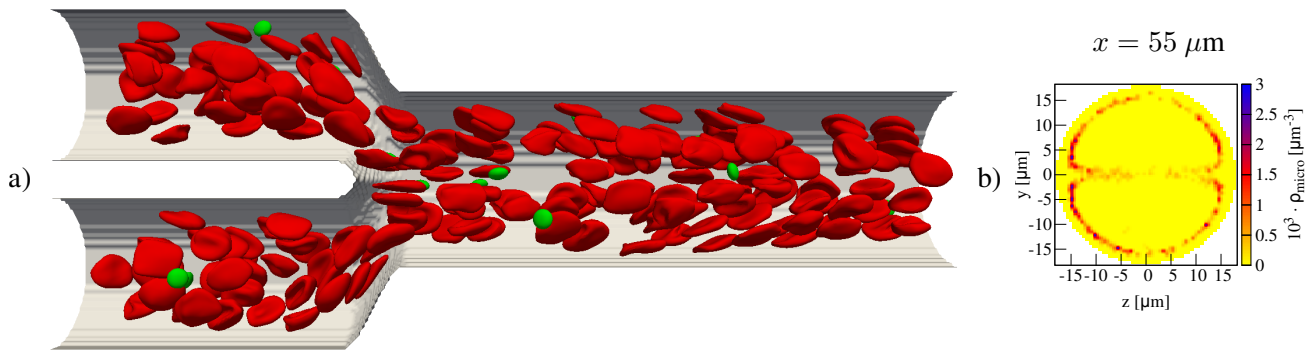

Fig. S6: (a) Simulation of microparticles with a platelet-like oblate shape flowing through a confluence. (b) Similar to the spherical microparticles of the main text, these particles also undergo anti-margination.

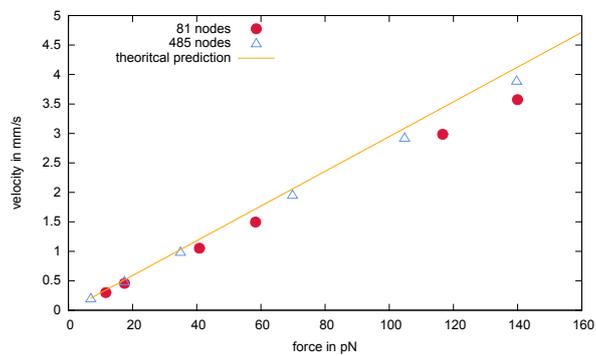

Fig. S7: Velocity of a spherical particle pulled through a fluid with a given force. Simulations with two different particle resolutions are compared to the theoretical prediction given by the Stokes drag.

increased resolution (485 nodes of the inner grid). Both show good agreement with the theoretical prediction and convergence to the theory for increasing grid resolution.

In order to prove mesh insensitivity we reduced the resolution of the membrane mesh of the red blood cells to 258 nodes and 512 triangles and the resolution of the stiff particles to 66 nodes and 128 triangles. In the same way the fluid mesh changes from 288x110x58 to 200x82x42. The red blood cell distribution behind a confluence and the cross-sectional microparticle concentration are compared to the results of the manuscript in figure S8. The results are in very good agreement. Small discrepancies may be caused by slightly different inflow concentrations.

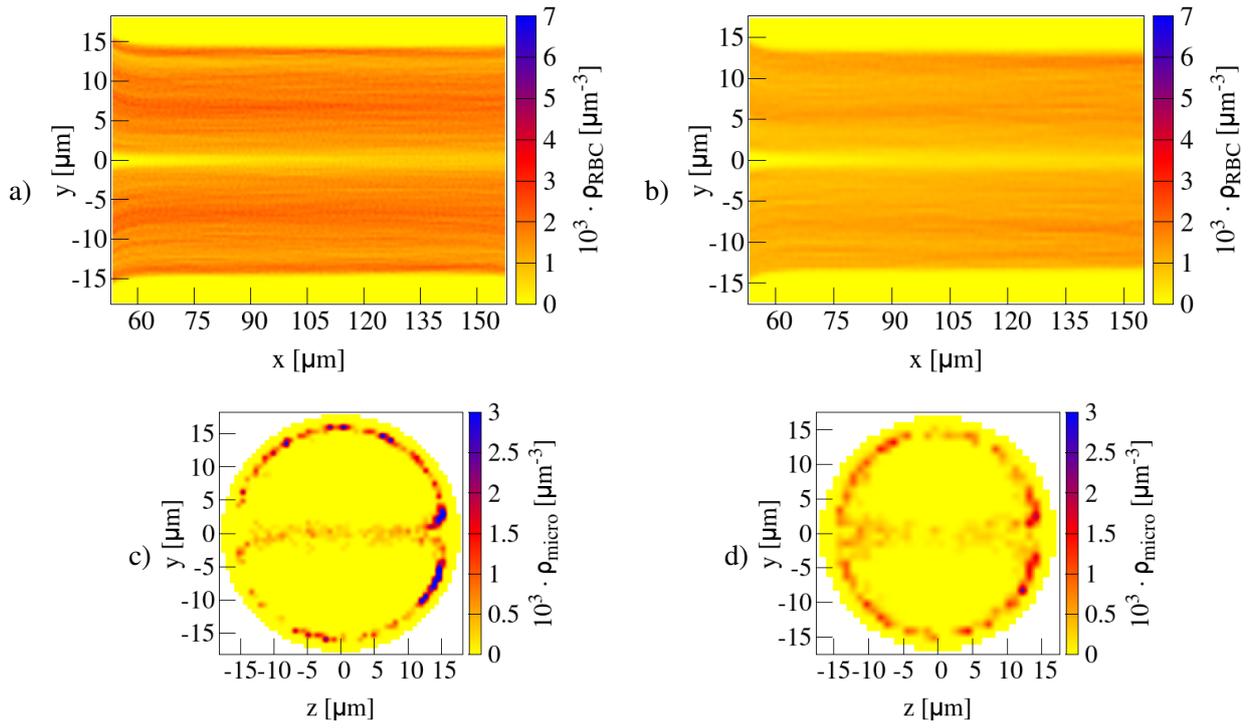

Fig. S8: The main results for different mesh resolutions: The cell free-layer in a), b) and the microparticle anti-margination in c), d). a),c) resolution used in the main text and b), d) decreased resolution (RBC: 512, microparticle: 128, fluid: 200x82x42). Both resolutions lead to similar results for central cell free layer stability and fraction of anti-marginated microparticles (c) 15.8% and d) 16.2%). Figures a) and c) are from the main text.



\end{document}